\documentclass[journal,twoside,twocolumn]{IEEEtran}

\usepackage{cite}
\usepackage[T1]{fontenc}
\usepackage{graphicx}
\usepackage{amssymb}
\usepackage{amsfonts}
\usepackage{amsmath}
\usepackage{amsthm}
\usepackage{mathrsfs}
\usepackage{xspace}
\usepackage{bm}
\usepackage{upgreek}

\usepackage{paralist}
\usepackage{microtype}
\usepackage{hyperref}
\usepackage{url}
\usepackage{balance}
\usepackage{subfigure}
\usepackage{xcolor}
\usepackage{fancyref}
\usepackage{framed}
\usepackage{booktabs}
\usepackage{nicefrac}
\usepackage{textcomp}
\usepackage{multirow}
\usepackage{stmaryrd}
 
\usepackage{algorithm}
\usepackage{algpseudocode}

\allowdisplaybreaks

\newtheorem{alg}{Algorithm}

\newcommand{\safemath}[2]{\newcommand{#1}{\ensuremath{#2}\xspace}}

\safemath{\bma}{\mathbf{a}}
\safemath{\bmb}{\mathbf{b}}
\safemath{\bmc}{\mathbf{c}}
\safemath{\bmd}{\mathbf{d}}
\safemath{\bme}{\mathbf{e}}
\safemath{\bmf}{\mathbf{f}}
\safemath{\bmg}{\mathbf{g}}
\safemath{\bmh}{\mathbf{h}}
\safemath{\bmi}{\mathbf{i}}
\safemath{\bmj}{\mathbf{j}}
\safemath{\bmk}{\mathbf{k}}
\safemath{\bml}{\mathbf{l}}
\safemath{\bmm}{\mathbf{m}}
\safemath{\bmn}{\mathbf{n}}
\safemath{\bmo}{\mathbf{o}}
\safemath{\bmp}{\mathbf{p}}
\safemath{\bmq}{\mathbf{q}}
\safemath{\bmr}{\mathbf{r}}
\safemath{\bms}{\mathbf{s}}
\safemath{\bmt}{\mathbf{t}}
\safemath{\bmu}{\mathbf{u}}
\safemath{\bmv}{\mathbf{v}}
\safemath{\bmw}{\mathbf{w}}
\safemath{\bmx}{\mathbf{x}}
\safemath{\bmy}{\mathbf{y}}
\safemath{\bmz}{\mathbf{z}}
\safemath{\bmzero}{\mathbf{0}}
\safemath{\bmone}{\mathbf{1}}

\bmdefine{\biad}{a}
\bmdefine{\bibd}{b}
\bmdefine{\bicd}{c}
\bmdefine{\bidd}{d}
\bmdefine{\bied}{e}
\bmdefine{\bifd}{f}
\bmdefine{\bigd}{g}
\bmdefine{\bihd}{h}
\bmdefine{\biid}{i}
\bmdefine{\bijd}{j}
\bmdefine{\bikd}{k}
\bmdefine{\bild}{l}
\bmdefine{\bimd}{m}
\bmdefine{\bind}{n}
\bmdefine{\biod}{o}
\bmdefine{\bipd}{p}
\bmdefine{\biqd}{q}
\bmdefine{\bird}{r}
\bmdefine{\bisd}{s}
\bmdefine{\bitd}{t}
\bmdefine{\biud}{u}
\bmdefine{\bivd}{v}
\bmdefine{\biwd}{w}
\bmdefine{\bixd}{x}
\bmdefine{\biyd}{y}
\bmdefine{\bizd}{z}

\bmdefine{\bixid}{\xi}
\bmdefine{\bilambdad}{\lambda}
\bmdefine{\bimud}{\mu}
\bmdefine{\bithetad}{\theta}
\bmdefine{\biphid}{\phi}
\bmdefine{\bideltad}{\delta}

\safemath{\bmia}{\biad}
\safemath{\bmib}{\bibd}
\safemath{\bmic}{\bicd}
\safemath{\bmid}{\bidd}
\safemath{\bmie}{\bied}
\safemath{\bmif}{\bifd}
\safemath{\bmig}{\bigd}
\safemath{\bmih}{\bihd}
\safemath{\bmii}{\biid}
\safemath{\bmij}{\bijd}
\safemath{\bmik}{\bikd}
\safemath{\bmil}{\bild}
\safemath{\bmim}{\bimd}
\safemath{\bmin}{\bind}
\safemath{\bmio}{\biod}
\safemath{\bmip}{\bipd}
\safemath{\bmiq}{\biqd}
\safemath{\bmir}{\bird}
\safemath{\bmis}{\bisd}
\safemath{\bmit}{\bitd}
\safemath{\bmiu}{\biud}
\safemath{\bmiv}{\bivd}
\safemath{\bmiw}{\biwd}
\safemath{\bmix}{\bixd}
\safemath{\bmiy}{\biyd}
\safemath{\bmiz}{\bizd}

\safemath{\bmxi}{\bixid}
\safemath{\bmlambda}{\bilambdad}
\safemath{\bmmu}{\bimud}
\safemath{\bmtheta}{\bithetad}
\safemath{\bmphi}{\biphid}
\safemath{\bmdelta}{\bideltad}

\safemath{\bA}{\mathbf{A}}
\safemath{\bB}{\mathbf{B}}
\safemath{\bC}{\mathbf{C}}
\safemath{\bD}{\mathbf{D}}
\safemath{\bE}{\mathbf{E}}
\safemath{\bF}{\mathbf{F}}
\safemath{\bG}{\mathbf{G}}
\safemath{\bH}{\mathbf{H}}
\safemath{\bI}{\mathbf{I}}
\safemath{\bJ}{\mathbf{J}}
\safemath{\bK}{\mathbf{K}}
\safemath{\bL}{\mathbf{L}}
\safemath{\bM}{\mathbf{M}}
\safemath{\bN}{\mathbf{N}}
\safemath{\bO}{\mathbf{O}}
\safemath{\bP}{\mathbf{P}}
\safemath{\bQ}{\mathbf{Q}}
\safemath{\bR}{\mathbf{R}}
\safemath{\bS}{\mathbf{S}}
\safemath{\bT}{\mathbf{T}}
\safemath{\bU}{\mathbf{U}}
\safemath{\bV}{\mathbf{V}}
\safemath{\bW}{\mathbf{W}}
\safemath{\bX}{\mathbf{X}}
\safemath{\bY}{\mathbf{Y}}
\safemath{\bZ}{\mathbf{Z}}

\safemath{\bZero}{\mathbf{0}}
\safemath{\bOne}{\mathbf{1}}
\safemath{\bDelta}{\mathbf{\Delta}}
\safemath{\bLambda}{\mathbf{\UpLambda}}
\safemath{\bPhi}{\mathbf{\Upphi}}
\safemath{\bSigma}{\mathbf{\Upsigma}}
\safemath{\bOmega}{\mathbf{\Upomega}}
\safemath{\bTheta}{\mathbf{\Uptheta}}

\bmdefine{\biAd}{A}
\bmdefine{\biBd}{B}
\bmdefine{\biCd}{C}
\bmdefine{\biDd}{D}
\bmdefine{\biEd}{E}
\bmdefine{\biFd}{F}
\bmdefine{\biGd}{G}
\bmdefine{\biHd}{H}
\bmdefine{\biId}{I}
\bmdefine{\biJd}{J}
\bmdefine{\biKd}{K}
\bmdefine{\biLd}{L}
\bmdefine{\biMd}{M}
\bmdefine{\biOd}{N}
\bmdefine{\biPd}{O}
\bmdefine{\biQd}{P}
\bmdefine{\biRd}{R}
\bmdefine{\biSd}{S}
\bmdefine{\biTd}{T}
\bmdefine{\biUd}{U}
\bmdefine{\biVd}{V}
\bmdefine{\biWd}{W}
\bmdefine{\biXd}{X}
\bmdefine{\biYd}{Y}
\bmdefine{\biZd}{Z}

\bmdefine{\biDelta}{\Delta}
\bmdefine{\biLambda}{\Lambda}
\bmdefine{\biPhi}{\Phi}
\bmdefine{\biSigma}{\Sigma}
\bmdefine{\biOmega}{\Omega}
\bmdefine{\biTheta}{\Theta}

\safemath{\bimA}{\biAd}
\safemath{\bimB}{\biBd}
\safemath{\bimC}{\biCd}
\safemath{\bimD}{\biDd}
\safemath{\bimE}{\biEd}
\safemath{\bimF}{\biFd}
\safemath{\bimG}{\biGd}
\safemath{\bimH}{\biHd}
\safemath{\bimI}{\biId}
\safemath{\bimJ}{\biJd}
\safemath{\bimK}{\biKd}
\safemath{\bimL}{\biLd}
\safemath{\bimM}{\biMd}
\safemath{\bimN}{\biNd}
\safemath{\bimO}{\biOd}
\safemath{\bimP}{\biPd}
\safemath{\bimQ}{\biQd}
\safemath{\bimR}{\biRd}
\safemath{\bimS}{\biSd}
\safemath{\bimT}{\biTd}
\safemath{\bimU}{\biUd}
\safemath{\bimV}{\biVd}
\safemath{\bimW}{\biWd}
\safemath{\bimX}{\biXd}
\safemath{\bimY}{\biYd}
\safemath{\bimZ}{\biZd}

\safemath{\bimDelta}{\biDelta}
\safemath{\bimLambda}{\biLambda}
\safemath{\bimPhi}{\biPhi}
\safemath{\bimSigma}{\biSigma}
\safemath{\bimOmega}{\biOmega}
\safemath{\bimTheta}{\biTheta}

\safemath{\setA}{\mathcal{A}}
\safemath{\setB}{\mathcal{B}}
\safemath{\setC}{\mathcal{C}}
\safemath{\setD}{\mathcal{D}}
\safemath{\setE}{\mathcal{E}}
\safemath{\setF}{\mathcal{F}}
\safemath{\setG}{\mathcal{G}}
\safemath{\setH}{\mathcal{H}}
\safemath{\setI}{\mathcal{I}}
\safemath{\setJ}{\mathcal{J}}
\safemath{\setK}{\mathcal{K}}
\safemath{\setL}{\mathcal{L}}
\safemath{\setM}{\mathcal{M}}
\safemath{\setN}{\mathcal{N}}
\safemath{\setO}{\mathcal{O}}
\safemath{\setP}{\mathcal{P}}
\safemath{\setQ}{\mathcal{Q}}
\safemath{\setR}{\mathcal{R}}
\safemath{\setS}{\mathcal{S}}
\safemath{\setT}{\mathcal{T}}
\safemath{\setU}{\mathcal{U}}
\safemath{\setV}{\mathcal{V}}
\safemath{\setW}{\mathcal{W}}
\safemath{\setX}{\mathcal{X}}
\safemath{\setY}{\mathcal{Y}}
\safemath{\setZ}{\mathcal{Z}}
\safemath{\emptySet}{\varnothing}

\safemath{\colA}{\mathscr{A}}
\safemath{\colB}{\mathscr{B}}
\safemath{\colC}{\mathscr{C}}
\safemath{\colD}{\mathscr{D}}
\safemath{\colE}{\mathscr{E}}
\safemath{\colF}{\mathscr{F}}
\safemath{\colG}{\mathscr{G}}
\safemath{\colH}{\mathscr{H}}
\safemath{\colI}{\mathscr{I}}
\safemath{\colJ}{\mathscr{J}}
\safemath{\colK}{\mathscr{K}}
\safemath{\colL}{\mathscr{L}}
\safemath{\colM}{\mathscr{M}}
\safemath{\colN}{\mathscr{N}}
\safemath{\colO}{\mathscr{O}}
\safemath{\colP}{\mathscr{P}}
\safemath{\colQ}{\mathscr{Q}}
\safemath{\colR}{\mathscr{R}}
\safemath{\colS}{\mathscr{S}}
\safemath{\colT}{\mathscr{T}}
\safemath{\colU}{\mathscr{U}}
\safemath{\colV}{\mathscr{V}}
\safemath{\colW}{\mathscr{W}}
\safemath{\colX}{\mathscr{X}}
\safemath{\colY}{\mathscr{Y}}
\safemath{\colZ}{\mathscr{Z}}

\safemath{\opA}{\mathbb{A}}
\safemath{\opB}{\mathbb{B}}
\safemath{\opC}{\mathbb{C}}
\safemath{\opD}{\mathbb{D}}
\safemath{\opE}{\mathbb{E}}
\safemath{\opF}{\mathbb{F}}
\safemath{\opG}{\mathbb{G}}
\safemath{\opH}{\mathbb{H}}
\safemath{\opI}{\mathbb{I}}
\safemath{\opJ}{\mathbb{J}}
\safemath{\opK}{\mathbb{K}}
\safemath{\opL}{\mathbb{L}}
\safemath{\opM}{\mathbb{M}}
\safemath{\opN}{\mathbb{N}}
\safemath{\opO}{\mathbb{O}}
\safemath{\opP}{\mathbb{P}}
\safemath{\opQ}{\mathbb{Q}}
\safemath{\opR}{\mathbb{R}}
\safemath{\opS}{\mathbb{S}}
\safemath{\opT}{\mathbb{T}}
\safemath{\opU}{\mathbb{U}}
\safemath{\opV}{\mathbb{V}}
\safemath{\opW}{\mathbb{W}}
\safemath{\opX}{\mathbb{X}}
\safemath{\opY}{\mathbb{Y}}
\safemath{\opZ}{\mathbb{Z}}
\safemath{\opZero}{\mathbb{O}}
\safemath{\identityop}{\opI}

\safemath{\veca}{\bma}
\safemath{\vecb}{\bmb}
\safemath{\vecc}{\bmc}
\safemath{\vecd}{\bmd}
\safemath{\vece}{\bme}
\safemath{\vecf}{\bmf}
\safemath{\vecg}{\bmg}
\safemath{\vech}{\bmh}
\safemath{\veci}{\bmi}
\safemath{\vecj}{\bmj}
\safemath{\veck}{\bmk}
\safemath{\vecl}{\bml}
\safemath{\vecm}{\bmm}
\safemath{\vecn}{\bmn}
\safemath{\veco}{\bmo}
\safemath{\vecp}{\bmp}
\safemath{\vecq}{\bmq}
\safemath{\vecr}{\bmr}
\safemath{\vecs}{\bms}
\safemath{\vect}{\bmt}
\safemath{\vecu}{\bmu}
\safemath{\vecv}{\bmv}
\safemath{\vecw}{\bmw}
\safemath{\vecx}{\bmx}
\safemath{\vecy}{\bmy}
\safemath{\vecz}{\bmz}

\safemath{\veczero}{\bmzero}
\safemath{\vecone}{\bmone}
\safemath{\vecxi}{\bmxi}
\safemath{\veclambda}{\bmlambda}
\safemath{\vecmu}{\bmmu}
\safemath{\vectheta}{\bmtheta}
\safemath{\vecphi}{\bmphi}
\safemath{\vecdelta}{\bmdelta}

\safemath{\matA}{\bA}
\safemath{\matB}{\bB}
\safemath{\matC}{\bC}
\safemath{\matD}{\bD}
\safemath{\matE}{\bE}
\safemath{\matF}{\bF}
\safemath{\matG}{\bG}
\safemath{\matH}{\bH}
\safemath{\matI}{\bI}
\safemath{\matJ}{\bJ}
\safemath{\matK}{\bK}
\safemath{\matL}{\bL}
\safemath{\matM}{\bM}
\safemath{\matN}{\bN}
\safemath{\matO}{\bO}
\safemath{\matP}{\bP}
\safemath{\matQ}{\bQ}
\safemath{\matR}{\bR}
\safemath{\matS}{\bS}
\safemath{\matT}{\bT}
\safemath{\matU}{\bU}
\safemath{\matV}{\bV}
\safemath{\matW}{\bW}
\safemath{\matX}{\bX}
\safemath{\matY}{\bY}
\safemath{\matZ}{\bZ}
\safemath{\matzero}{\bmzero}

\safemath{\matDelta}{\bDelta}
\safemath{\matLambda}{\bLambda}
\safemath{\matPhi}{\bPhi}
\safemath{\matSigma}{\bSigma}
\safemath{\matOmega}{\bOmega}
\safemath{\matTheta}{\bTheta}

\safemath{\matidentity}{\matI}
\safemath{\matone}{\matO}

\safemath{\rnda}{A}
\safemath{\rndb}{B}
\safemath{\rndc}{C}
\safemath{\rndd}{D}
\safemath{\rnde}{E}
\safemath{\rndf}{F}
\safemath{\rndg}{G}
\safemath{\rndh}{H}
\safemath{\rndi}{I}
\safemath{\rndj}{J}
\safemath{\rndk}{K}
\safemath{\rndl}{L}
\safemath{\rndm}{M}
\safemath{\rndn}{N}
\safemath{\rndo}{O}
\safemath{\rndp}{P}
\safemath{\rndq}{Q}
\safemath{\rndr}{R}
\safemath{\rnds}{S}
\safemath{\rndt}{T}
\safemath{\rndu}{U}
\safemath{\rndv}{V}
\safemath{\rndw}{W}
\safemath{\rndx}{X}
\safemath{\rndy}{Y}
\safemath{\rndz}{Z}

\safemath{\rveca}{\bimA}
\safemath{\rvecb}{\bimB}
\safemath{\rvecc}{\bimC}
\safemath{\rvecd}{\bimD}
\safemath{\rvece}{\bimE}
\safemath{\rvecf}{\bimF}
\safemath{\rvecg}{\bimG}
\safemath{\rvech}{\bimH}
\safemath{\rveci}{\bimI}
\safemath{\rvecj}{\bimJ}
\safemath{\rveck}{\bimK}
\safemath{\rvecl}{\bimL}
\safemath{\rvecm}{\bimM}
\safemath{\rvecn}{\bimN}
\safemath{\rveco}{\bomO}
\safemath{\rvecp}{\bimP}
\safemath{\rvecq}{\bimQ}
\safemath{\rvecr}{\bimR}
\safemath{\rvecs}{\bimS}
\safemath{\rvect}{\bimT}
\safemath{\rvecu}{\bimU}
\safemath{\rvecv}{\bimV}
\safemath{\rvecw}{\bimW}
\safemath{\rvecx}{\bimX}
\safemath{\rvecy}{\bimY}
\safemath{\rvecz}{\bimZ}

\safemath{\rvecxi}{\bmxi}
\safemath{\rveclambda}{\bmlambda}
\safemath{\rvecmu}{\bmmu}
\safemath{\rvectheta}{\bmtheta}
\safemath{\rvecphi}{\bmphi}

\safemath{\rmatA}{\bimA}
\safemath{\rmatB}{\bimB}
\safemath{\rmatC}{\bimC}
\safemath{\rmatD}{\bimD}
\safemath{\rmatE}{\bimE}
\safemath{\rmatF}{\bimF}
\safemath{\rmatG}{\bimG}
\safemath{\rmatH}{\bimH}
\safemath{\rmatI}{\bimI}
\safemath{\rmatJ}{\bimJ}
\safemath{\rmatK}{\bimK}
\safemath{\rmatL}{\bimL}
\safemath{\rmatM}{\bimM}
\safemath{\rmatN}{\bimN}
\safemath{\rmatO}{\bimO}
\safemath{\rmatP}{\bimP}
\safemath{\rmatQ}{\bimQ}
\safemath{\rmatR}{\bimR}
\safemath{\rmatS}{\bimS}
\safemath{\rmatT}{\bimT}
\safemath{\rmatU}{\bimU}
\safemath{\rmatV}{\bimV}
\safemath{\rmatW}{\bimW}
\safemath{\rmatX}{\bimX}
\safemath{\rmatY}{\bimY}
\safemath{\rmatZ}{\bimZ}

\safemath{\rmatDelta}{\bimDelta}
\safemath{\rmatLambda}{\bimLambda}
\safemath{\rmatPhi}{\bimPhi}
\safemath{\rmatSigma}{\bimSigma}
\safemath{\rmatOmega}{\bimOmega}
\safemath{\rmatTheta}{\bimTheta}

\newenvironment{textbmatrix}{	\setlength{\arraycolsep}{2.5pt}%
								\big[\begin{matrix}}{\end{matrix}\big]%
								\raisebox{0.08ex}{\vphantom{M}}}

\def\be{\begin{equation}}
\def\ee{\end{equation}}
\def\een{\nonumber \end{equation}}
\def\mat{\begin{bmatrix}}
\def\emat{\end{bmatrix}}
\def\btm{\begin{textbmatrix}}
\def\etm{\end{textbmatrix}}

\def\ba#1\ea{\begin{align}#1\end{align}}
\def\bas#1\eas{\begin{align*}#1\end{align*}}
\def\bs#1\es{\begin{split}#1\end{split}} 
\def\bg#1\eg{\begin{gather}#1\end{gather}}
\def\bml#1\eml{\begin{multline}#1\end{multline}}
\def\bi#1\ei{\begin{itemize}#1\end{itemize}}

\newcommand{\lefto}{\mathopen{}\left}

\DeclareMathOperator{\tr}{tr}				
\DeclareMathOperator*{\argmin}{arg\;min}		
\DeclareMathOperator{\Exop}{\opE}			

\newcommand{\Ex}[2]{\ensuremath{\Exop_{#1}\lefto[#2\right]}} 	

\newcommand{\vecnorm}[1]{\lefto\lVert#1\right\rVert}		

\safemath{\dirac}{\delta}					
\safemath{\krond}{\dirac}					

\safemath{\upto}{\uparrow}
\safemath{\downto}{\downarrow}
\safemath{\iu}{j}							
\safemath{\ev}{\lambda}						
\safemath{\hilseqspace}{l^{2}}				
\newcommand{\banachfunspace}[1]{\setL^{#1}}	
\safemath{\hilfunspace}{\banachfunspace{2}}	

\safemath{\SNR}{\textsf{SNR}} 				
\safemath{\PAR}{\textsf{PAR}} 				
\safemath{\No}{N_0}							
\safemath{\Es}{E_s}							
\safemath{\Eb}{E_b}							
\safemath{\EbNo}{\frac{\Eb}{\No}}
\safemath{\EsNo}{\frac{\Es}{\No}}

\DeclareMathOperator{\CHop}{\ensuremath{\opH}} 
\safemath{\tvir}{\rndh_{\CHop}}				
\safemath{\tvtf}{\rndl_{\CHop}}				
\safemath{\spf}{\rnds_{\CHop}}				
\safemath{\bff}{H_{\CHop}}					

\safemath{\ircf}{r_{h}}						
\safemath{\tftvcf}{r_{s}}					
\safemath{\tfcf}{r_{l}}						
\safemath{\bfcf}{r_{H}}						

\safemath{\tcorr}{c_h}						
\safemath{\scf}{c_{s}}						
\safemath{\tfcorr}{c_{l}}					
\safemath{\fcorr}{c_{H}}						

\safemath{\mi}{I}							
\safemath{\capacity}{C}						

\safemath{\normal}{\mathcal{N}}			
\safemath{\jpg}{\mathcal{CN}}			
\safemath{\mchain}{\leftrightarrow}		

\safemath{\dB}{\,\mathrm{dB}}
\safemath{\dBm}{\,\mathrm{dBm}}
\safemath{\Hz}{\,\mathrm{Hz}}
\safemath{\kHz}{\,\mathrm{kHz}}
\safemath{\MHz}{\,\mathrm{MHz}}
\safemath{\GHz}{\,\mathrm{GHz}}
\safemath{\s}{\,\mathrm{s}}
\safemath{\ms}{\,\mathrm{ms}}
\safemath{\mus}{\,\mathrm{\text{\textmu}s}}
\safemath{\ns}{\,\mathrm{ns}}
\safemath{\ps}{\,\mathrm{ps}}
\safemath{\meter}{\,\mathrm{m}}
\safemath{\mm}{\,\mathrm{mm}}
\safemath{\cm}{\,\mathrm{cm}}
\safemath{\m}{\,\mathrm{m}}
\safemath{\W}{\,\mathrm{W}}
\safemath{\mW}{\, \mathrm{mW}}
\safemath{\J}{\,\mathrm{J}}
\safemath{\K}{\,\mathrm{K}}
\safemath{\bit}{\,\mathrm{bit}}
\safemath{\nat}{\,\mathrm{nat}}

\safemath{\define}{\triangleq}			

\safemath{\equivalent}{\sim}
\safemath{\distas}{\sim}					
\safemath{\sdiff}{\Delta}				

\safemath{\reals}{\mathbb{R}}
\safemath{\positivereals}{\reals_{+}}
\safemath{\integers}{\mathbb{Z}}
\safemath{\posint}{\integers_{+}}
\safemath{\naturals}{\mathbb{N}}
\safemath{\posnaturals}{\naturals_{+}}
\safemath{\complexset}{\mathbb{C}}
\safemath{\rationals}{\mathbb{Q}}

\newcommand*{\fancyrefapplabelprefix}{app}		
\newcommand*{\fancyrefthmlabelprefix}{thm}		
\newcommand*{\fancyreflemlabelprefix}{lem}		
\newcommand*{\fancyrefcorlabelprefix}{cor}		
\newcommand*{\fancyrefdeflabelprefix}{def}		
\newcommand*{\fancyrefproplabelprefix}{prop}	
\newcommand*{\fancyrefobslabelprefix}{obs}		
\newcommand*{\fancyrefalglabelprefix}{alg}		
\newcommand*{\fancyrefasmlabelprefix}{asm}	    
\newcommand*{\fancyreftbllabelprefix}{tbl}	    

\frefformat{vario}{\fancyrefseclabelprefix}{Section~#1}
\frefformat{vario}{\fancyrefthmlabelprefix}{Theorem~#1}
\frefformat{vario}{\fancyreflemlabelprefix}{Lemma~#1}
\frefformat{vario}{\fancyrefcorlabelprefix}{Corollary~#1}
\frefformat{vario}{\fancyrefdeflabelprefix}{Definition~#1}
\frefformat{vario}{\fancyrefobslabelprefix}{Observation~#1}
\frefformat{vario}{\fancyrefasmlabelprefix}{Assumption~#1}
\frefformat{vario}{\fancyreffiglabelprefix}{Figure~#1}
\frefformat{vario}{\fancyrefapplabelprefix}{Appendix~#1} 
\frefformat{vario}{\fancyrefproplabelprefix}{Proposition~#1}
\frefformat{vario}{\fancyrefalglabelprefix}{Algorithm~#1}
\frefformat{vario}{\fancyrefeqlabelprefix}{(#1)}
\frefformat{vario}{\fancyreftbllabelprefix}{Table~#1}

\newtheorem{thm}{Theorem}

\newtheorem{defi}{Definition}
\newtheorem{lem}[thm]{Lemma} 

\newtheorem{rem}{Remark}

\safemath{\dictab}{[\,\dicta\,\,\dictb\,]}

\safemath{\ysig}{\bmy}
\safemath{\ysighat}{\hat{\ysig}}
\safemath{\ysigdim}{M}
\safemath{\xsig}{\bmx}
\safemath{\xsigdim}{N}
\safemath{\nx}{n_x}
\safemath{\zsig}{\bmz}
\safemath{\zsigdim}{\ysigdim}
\safemath{\rsig}{\bmr}
\safemath{\Adict}{\bA}
\safemath{\Adicttilde}{\widetilde{\Adict}}
\safemath{\Adictdim}{\outputdim\times\xsigdim}
\safemath{\avec}{\bma}
\safemath{\avectilde}{\tilde{\avec}}
\safemath{\Bdict}{\bB}
\safemath{\Bdicttilde}{\widetilde{\Bdict}}
\safemath{\Cdict}{\bC}
\safemath{\cvec}{\bmc}
\safemath{\Ddict}{\bD}
\safemath{\Ddictdim}{\ysigdim\times\xsigdim}
\safemath{\dvec}{\bmd}
\safemath{\Ddicttilde}{\widetilde{\bD}}
\safemath{\Bonb}{\bB}
\safemath{\bvec}{\bmb}
\safemath{\Bonbdim}{\ysigdim\times\ysigdim}
\safemath{\noise}{\bmn}
\safemath{\noisedim}{\ysigim}
\safemath{\err}{\bme}
\safemath{\errdim}{\ysigdim}
\safemath{\errset}{\setE}
\safemath{\nerr}{n_e}
\safemath{\delop}{\bP_\errset}
\safemath{\delopc}{\bP_{{\errset}^c}}

\safemath{\cplxi}{\imath}
\safemath{\cplxj}{\jmath}

\safemath{\dict}{\matD}
\safemath{\inputdim}{N}		
\safemath{\outputdim}{M}		
\safemath{\sparsity}{S}	
\safemath{\inputdimA}{{N_a}}	
\safemath{\inputdimB}{{N_b}}	
\safemath{\elemA}{{n_a}}	
\safemath{\elemB}{{n_b}}	
\safemath{\resA}{\matR_a}	
\safemath{\resB}{\matR_b}	
\safemath{\subD}{\matS} 
\safemath{\subA}{\matS_a} 
\safemath{\subB}{\matS_b} 
\safemath{\dicta}{\matA} 	
\safemath{\dictb}{\matB} 	
\safemath{\hollowS}{H}
\safemath{\hollowA}{H_a}
\safemath{\hollowB}{H_b}
\safemath{\cross}{Z}
\safemath{\coh}{\mu_d}			
\safemath{\coha}{\mu_a}			
\safemath{\cohb}{\mu_b}			
\safemath{\mubs}{\nu}	
\safemath{\cohm}{\mu_m} 
\safemath{\dictset}{\setD}	
\safemath{\dictsetp}{\dictset(\coh,\coha,\cohb)}	
\safemath{\dictsetgen}{\dictset_\text{gen}}
\safemath{\dictsetgenp}{\dictsetgen(\coh)}
\safemath{\dictsetonb}{\dictset_\text{onb}}
\safemath{\dictsetonbp}{\dictsetonb(\coh)}

\safemath{\leftside}{U}
\safemath{\rightsideA}{R_a}
\safemath{\rightsideB}{R_b}

\safemath{\indexS}{\setI_S} 

\safemath{\na}{n_a}			
\safemath{\nb}{n_b}			
\safemath{\coeffa}{p_i}	
\safemath{\coeffb}{q_j}	
\safemath{\seta}{\setP}		
\safemath{\setb}{\setQ}     
\safemath{\setw}{\setW}	
\safemath{\setz}{\setZ}	
\safemath{\cola}{\veca}		
\safemath{\colb}{\vecb}		
\safemath{\cold}{\vecd}		
\safemath{\inputvec}{\vecx} 	
\safemath{\error}{\vece}	
\safemath{\noiseout}{\vecz} 	
\safemath{\inputvecel}{x}
\safemath{\inputveca}{\vecx_a}
\safemath{\inputvecb}{\vecx_b}
\safemath{\outputvec}{\vecy}	
\safemath{\lambdamin}{\lambda_{\mathrm{min}}}

\safemath{\elltwo}{\ell_2}
\safemath{\ellone}{\ell_1}
\safemath{\ellzero}{\ell_0}
\safemath{\ellinf}{\ell_\infty}
\safemath{\ellinftilde}{\ell_{\widetilde\infty}}
\safemath{\licard}{Z(\coh,\coha,\cohb)}
\safemath{\xsol}{\hat{x}}
\safemath{\xbord}{x_b}		
\safemath{\xstat}{x_s}		
\safemath{\xstatLone}{\tilde{x}_s}
\safemath{\order}{\mathcal{O}} 
\safemath{\scales}{\Theta} 
\safemath{\ones}{\mathbf{1}} 
\safemath{\zeroes}{\mathbf{0}} 
\safemath{\thlone}{\kappa(\coh,\cohb)} 
\safemath{\constoneA}{\delta} 
\safemath{\constoneB}{\epsilon} 
\safemath{\nlarge}{L}				   
\safemath{\sumlarge}{S_\nlarge}
\safemath{\maxlarger}{P_\nlarge}	   
\safemath{\Pzero}{\textrm{P0}}	
\safemath{\Pone}{\textrm{P1}}
\safemath{\vecfir}{\vecw}			 
\safemath{\vecsec}{\vecz}
\safemath{\elvecfir}{w}              
\safemath{\elvecsec}{z}				 
\safemath{\nlargefir}{n}
\safemath{\normout}{\gamma}
\safemath{\auxfun}{h}
\safemath{\supp}{\textrm{supp}}

\safemath{\indexa}{\ell}
\safemath{\indexb}{r}
\safemath{\indexc}{i}
\safemath{\indexd}{j}

\safemath{\project}{P}

\begin{document}

\title{Finite-Alphabet MMSE Equalization for All-Digital Massive MU-MIMO mmWave Communication}
\author{Oscar Casta\~neda, Sven Jacobsson, Giuseppe Durisi, Tom Goldstein, and Christoph Studer\thanks{O.~Casta\~neda  and C.~Studer are with the School of Electrical and Computer Engineering, Cornell University, Ithaca, NY 14853 USA and with Cornell Tech, New York, NY 10044 USA (e-mail: {oc66@cornell.edu}; {studer@cornell.edu}).}\thanks{S.\ Jacobsson is with Ericsson Research, 417 56 Gothenburg, Sweden, and with Chalmers University of Technology, 412 58 Gothenburg, Sweden (e-mail: {sven.jacobsson@ericsson.com}).}\thanks{G.\ Durisi is with Chalmers University of Technology, 412 58 Gothenburg, Sweden (e-mail: {durisi@chalmers.se}).}\thanks{T. Goldstein is with the Department of Computer Science, University of Maryland, College Park, MD 20742 USA (e-mail: {tomg@cs.umd.edu}).}\thanks{The work of OC was supported in part by ComSenTer, one of six centers in JUMP, a Semiconductor Research Corporation (SRC) program sponsored by DARPA, by SRC nCORE task 2758.004, and by a Qualcomm Innovation Fellowship. The work of SJ and GD was supported by the Swedish Foundation for Strategic Research under grants SM13-0028 and ID14-0022, and by the Swedish Governmental Agency for Innovation Systems (VINNOVA) within the VINN Excellence center Chase. The work of TG was supported in part by the US National Science Foundation (NSF) under grant CCF-1535902 and by the US Office of Naval Research under grant N00014-15-1-2676. The work of CS was supported in part by Xilinx Inc.\ and by the US NSF under grants ECCS-1408006 and CCF-1535897.}\thanks{A MATLAB simulator for the FAME approach proposed in this paper is available on GitHub: \url{https://github.com/quantizedmassivemimo/fame}}
}

\maketitle


\begin{abstract}
We propose finite-alphabet equalization, a new paradigm that restricts the entries of the spatial equalization matrix to low-resolution numbers, enabling  high-throughput, low-power, and low-cost hardware equalizers.
To minimize the performance loss of this paradigm, we introduce FAME, short for finite-alphabet minimum mean-square error (MMSE) equalization, which is able to significantly outperform a na\"ive quantization of the linear MMSE matrix. 
We develop efficient algorithms to approximately solve the NP-hard FAME problem and showcase that near-optimal performance can be achieved with equalization coefficients quantized to only 1--3 bits for massive multi-user multiple-input multiple-output (MU-MIMO) millimeter-wave (mmWave) systems.
We provide very-large scale integration (VLSI) results that demonstrate a reduction in equalization power and area by at least a factor of $\bold{3.9}\boldsymbol{\times}$ and $\bold{5.8}\boldsymbol{\times}$, respectively.
\end{abstract}


\section{Introduction}

\IEEEPARstart{F}{uture} wireless systems are expected to deliver even higher data-rates within the already crowded frequency spectrum.
Emerging technologies, such as millimeter-wave (mmWave) communication \cite{mmWillWork,SwindlehurstCommMag} and massive multi-user multiple-input multiple-output (MU-MIMO) \cite{RPLLETM12,larsson14-02a}, have risen as promising candidates to provide such high data-rates.
The abundance of available bandwidth at mmWave frequencies, combined with the fine-grained beamforming capabilities provided by massive MU-MIMO, enables high-throughput communication to multiple user equipments (UEs) in the same time-frequency resource.
However, the presence of hundreds of antennas at the basestation (BS), each receiving  a wideband signal, necessitates sophisticated radio frequency (RF) and digital baseband processing circuitry.
As a result, circuit power consumption and system costs may increase significantly, which may hamper the success of this technology.

To reduce power consumption, the literature has largely focused on multi-antenna mmWave architectures that rely on hybrid analog-digital solutions \cite{roh14,sadhu17,alkhateeb14b}.
Albeit energy efficient, such architectures have limited multiplexing capabilities as they are only capable of simultaneously combining signals coming from a restricted number of directions \cite{pi2011,alkhateeb14b,bjornson19mimommwave,dutta19}; this key limitation may result in a reduced spectral efficiency. 
An emerging alternative is the use of all-digital BS architectures~\cite{mo15CAOQ,Roth17Digital,jacobssonTAMMU17}.
While it is commonly believed that all-digital BS designs would be energy inefficient, it has been shown recently that the power consumption of the RF and data-conversion elements in an all-digital BS is comparable to that of hybrid solutions, provided that the resolution of the data converters at the BS is suitably reduced~\cite{Roth17Digital,dutta19}.
The power consumption and system costs of baseband processing for all-digital BS architectures is, however, largely unexplored.

\subsection{The Case for Efficient Spatial Equalization}
\label{sec:intro_example}
Spatial equalization in the uplink (UEs transmit to BS) is among the most power- and throughput-critical tasks in all-digital BS architectures.
The purpose of spatial equalization is to collect the signals from all $U$ UEs at the $B$ BS antennas, while suppressing inter-UE interference.
Mathematically, spatial equalization amounts to one or multiple $U\times B$ matrix-vector multiplications involving a $U\times B$ equalization matrix and the $B$-dimensional received vector. 
These multiplications need to be performed on a per-baseband-sample basis (at the sample rate of the analog-to-digital converters).
Unfortunately, for a BS with $B=256$ antennas serving $U=16$ UEs, a conventional matrix-vector-product circuit operating at $2$\,G\,vectors/s consumes over $28$\,W and occupies more than $128\,\text{mm}^2$ when implemented in 28\,nm CMOS (see \fref{sec:vlsi} for the details). 
If more BS antennas and/or more UEs are considered, circuit power and area increase even further.
Clearly, more efficient spatial equalization circuitry is necessary for all-digital BS architectures operating at mmWave frequencies in order to minimize power consumption and silicon area (which translate to system costs), while achieving  high spectral efficiency. 

The matrix-vector products required for spatial equalization involve multiplications and additions, where the hardware multipliers dominate power and area. 
The area and delay of a hardware multiplier scales with $O(mn)$ and $O(\log(\max\{m,n\}))$, respectively, where $m$ and $n$ are the number of bits of each operand \cite{Zimmerman99}.
As a consequence, circuit area, delay, and power consumption (which is roughly proportional to circuit area) of a matrix-vector-product engine can be minimized by using a low number of bits to represent both operands.
The literature has extensively explored the efficacy of low-resolution data converters at the BS antennas of massive MU-MIMO systems~\cite{mo15CAOQ,alkhateeb14b,studer16a,Roth17Digital,mo17b,yan19,dutta19}. Depending on the scenario, $3$ to $8$ bits were shown to achieve near-optimal spectral efficiency~\cite{studer16a,Roth17Digital,mo17b,yan19,dutta19}.
Such methods reduce the precision of one of the operands (i.e., that of the received vector) in a matrix-vector product.
However, the coefficients of the equalization matrix (the other operand) are typically left at relatively high precision, e.g., $10$ to $12$ bits \cite{studer2011asic,WYWDCS2014}.
\begin{figure}[tp]
\centering
\includegraphics[width=0.85\columnwidth]{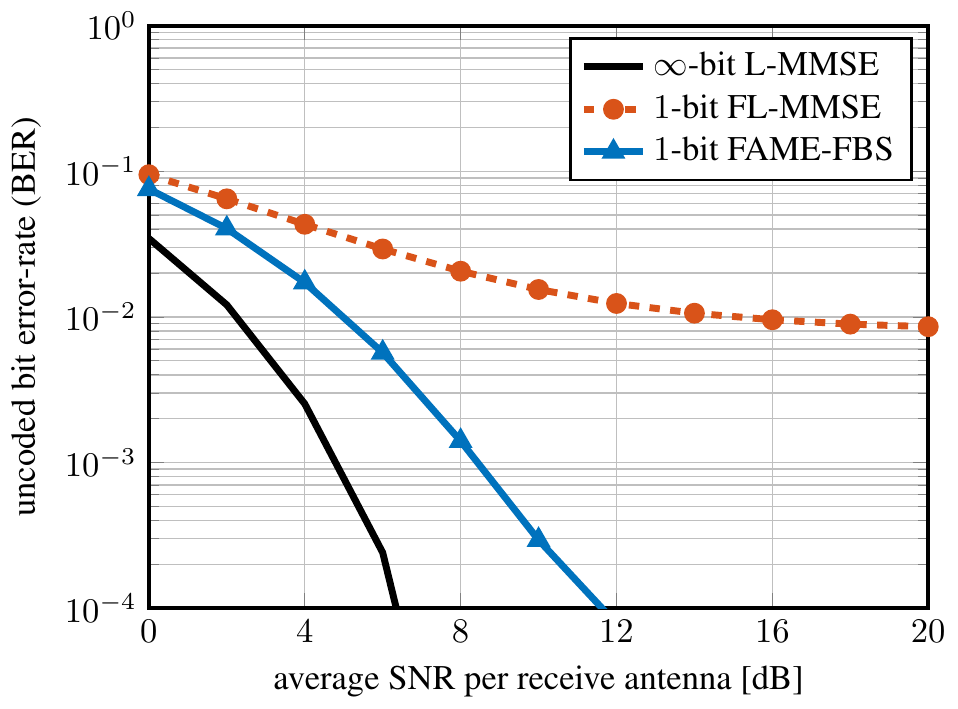}
\caption{Uncoded bit error-rate (BER) for a $B=256$ BS antenna, $U=16$ UE massive MU-MIMO system with $16$-QAM in an i.i.d.\ Rayleigh-fading channel. The FAME-FBS algorithm proposed in this paper significantly outperforms the na\"ive FL-MMSE equalizer, which quantizes  the real and imaginary parts  of the conventional, high-resolution L-MMSE equalizer to $1$ bit.}
\label{fig:ber_example}
\end{figure}
\subsection{Contributions}
To  reduce power consumption and implementation costs of spatial equalization, we propose to \emph{coarsely} quantize the coefficients of spatial equalization matrices, a paradigm that we call \emph{finite-alphabet equalization}.
We emphasize that, in contrast to approaches that use low-resolution analog-to-digital converters (ADCs) to quantize the received vector to be equalized \cite{mo15CAOQ,alkhateeb14b,studer16a,Roth17Digital,mo17b,yan19,dutta19}, finite-alphabet equalization intends to coarsely quantize the entries of the \emph{spatial equalization matrix}.
Although a straightforward concept, it turns out that obtaining low-resolution finite-alphabet equalization matrices that achieve high spectral efficiency is a hard problem.
\fref{fig:ber_example} illustrates this claim for a case where the coefficients of a spatial equalization matrix are quantized using $1$ bit per real and imaginary part.
A na\"ive quantization of the linear minimum mean-square error (L-MMSE) equalization matrix to $1$-bit leads to a finite-alphabet L-MMSE (FL-MMSE) equalizer, which, as we can see from \fref{fig:ber_example}, suffers from a high error floor.
To combat this problem, we propose \emph{finite-alphabet minimum mean-square error equalization} (FAME), which leads to an NP-hard optimization problem that can be solved approximately (and efficiently) using forward-backward splitting (FBS).
We refer to the resulting method as FAME-FBS. 
As shown in \fref{fig:ber_example}, using FAME-FBS results in a substantially improved error rate compared to FL-MMSE equalization.

The main contributions of this paper are summarized as follows.
We propose a specific finite-alphabet equalization-matrix structure that enables one to reduce the complexity of a $U\times B$ matrix-vector product by using $U\times B$ low-resolution coefficients, while still being able to deliver a performance similar to conventional, high-resolution spatial equalization matrices.
We derive the so-called FAME problem, whose solution leads to finite-alphabet equalization matrices that minimize the post-equalization mean-square error (MSE).
We propose a range of algorithms that approximate the NP-hard FAME problem---some of these algorithms achieve excellent performance even for 1-bit resolution; some require very low complexity. 
We present simulation results for line-of-sight (LoS) and non-LoS mmWave channel models, which demonstrate the efficacy of FAME in terms of error-vector magnitude (EVM), beamforming capabilities, and uncoded bit error-rate (BER). 
Finally, we implement reference finite-alphabet equalization circuits for different number of bits in 28\,nm CMOS to demonstrate the effectiveness of FAME in practice. 

\subsection{Notation}
Matrices and column vectors are represented with uppercase and lowercase boldface letters, respectively. 
The Hermitian transpose and the Frobenius norm of a matrix $\bA$ are denoted by $\bA^H$ and $\|\bA\|_{F}$, respectively.
The real part of a complex-valued matrix~$\bA$ is $\Re\{\bA\}$ and the imaginary part is $\Im\{\bA\}$. 
The $M\times M$ identity matrix is denoted by  $\bI_M$.
The $k$th entry and the $\ell_2$-norm of a vector~$\bma$ are $a_k$ and $\vecnorm{\veca}_2$, respectively; the entry-wise complex conjugate is denoted by $\bma^*$. 
The $k$th standard basis vector is represented by $\bme_k$.
The signum function~$\text{sgn}(\cdot)$ is defined as~$\text{sgn}(a)=+1$ for~$a\ge0$ and~$\text{sgn}(a)=-1$ for~$a<0$ and is applied entry-wise to vectors.
We use~$\Ex{\bmx}{\cdot}$ to denote expectation with respect to the random vector $\bmx$.
The set $\opS_+$ contains all positive semidefinite matrices, and the set $\reals_+$ contains all the non-negative real numbers.

\subsection{Paper Outline}
The rest of the paper is organized as follows. 
\fref{sec:systemmodel} introduces the system model and summarizes the basics of \mbox{L-MMSE} equalization.
\fref{sec:FAME} proposes the FAME problem and presents numerical experiments. 
\fref{sec:lowBitFAME} develops low-complexity algorithms that approximate the NP-hard FAME problem.
\fref{sec:vlsi} shows hardware implementation results.
We conclude in \fref{sec:conclusions}. Proofs and complexity counts are relegated to the appendices. 
\section{System Model and L-MMSE Equalization}
\label{sec:systemmodel}
\begin{figure}[tp]
\centering
\includegraphics[width=0.9\columnwidth]{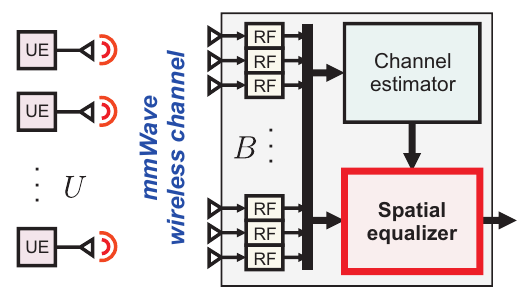}
\caption{Uplink of a massive MU-MIMO mmWave wireless communication system. The $U$ UEs transmit data in the same time-frequency resource to the $B$-antenna BS. After estimating the channel, the all-digital BS applies spatial equalization to collect the signals from the individual UEs and suppress inter-UE interference. In this work, we propose to use low-resolution spatial equalization matrices, an approach that we call finite-alphabet equalization.}
\label{fig:system_overview}
\end{figure}
We now introduce the system model considered in this work and briefly introduce the essentials of L-MMSE equalization. 
\subsection{System Model}
We focus on the uplink of a massive MU-MIMO system with $B$ BS antennas and $U\leq B$ single-antenna UEs, as illustrated in \fref{fig:system_overview}.
We consider the following narrowband  input-output relation:
\begin{align} \label{eq:systemmodel}
\bmy=\bH\bms+\bmn.
\end{align}
Here, $\bmy\in\complexset^B$ is the received signal vector at the BS,  $\bH\in\complexset^{B\times U}$ is the known uplink MIMO channel matrix, $\bms\in\setS^U$ is the transmit data vector, where $\setS$ is the constellation set (e.g., 16-QAM),  and  $\bmn\in\complexset^B$ is i.i.d.\ circularly-symmetric complex Gaussian noise with covariance matrix $\bC_\bmn=\Ex{\bmn}{\bmn\bmn^H}=\No\bI_B$ per complex entry. 
In what follows, we assume that the transmit signals of the UEs, $s_u$, $u=1,\ldots,U$, are i.i.d.\ zero mean with variance $\Es$ so that $\bC_\bms = \Ex{\bms}{\bms\bms^H}=\Es\bI_U$.

\begin{rem}
The input-output relation \fref{eq:systemmodel} is not only valid to model narrowband transmission, but can also be used to model the subcarriers of a wideband massive MU-MIMO system that uses orthogonal frequency-division multiplexing (OFDM). 
The theory and algorithms developed in the remainder of the paper can be generalized for systems with inter-symbol interference---the details are left for future work.
\end{rem}

\begin{rem}
We assume that the channel remains constant over multiple symbol transmissions and, hence, can be estimated.
For our mathematical derivations, we assume (quantized) perfect channel state information at the BS.
For systems in which the UEs use an antenna array to perform transmit beamforming, the channel matrix $\bH$  represents the joint effect of beamforming and the physical channel.
\end{rem}

\subsection{A Primer on L-MMSE Equalization}
\label{sec:lmmse}
A key task of the BS is to form estimates $\hat\bms\in\complexset^U$ of the transmitted data vector $\bms$. 
To develop methods that are computationally efficient and hardware friendly, we focus on linear spatial estimators of the form $\hat\bms = \bW^H \bmy$, 
where  $\bW^H\in\complexset^{U\times B}$ is the L-MMSE equalization matrix that minimizes the post-equalization MSE defined as 
\begin{align} \label{eq:MSE}
\textit{MSE} = \Ex{\bms,\bmn}{\|\bW^H\bmy-\bms\|^2_2}\!.
\end{align}
Given the assumptions on the statistics of the transmit data and noise vectors, $\bms$ and $\bmn$, we have that
\begin{align}
\textit{MSE} 
& =   \Es\|\bW^H\bH-\bI_U\|^2_F + \No\|\bW^H\|^2_F.
\end{align}
Hence, the L-MMSE equalization matrix can be obtained by solving the following matrix least-squares problem:  
\begin{align}
\label{eq:LMMSEmatrixProblem}
\bW^H  = \argmin_{\tilde\bW^H\in\complexset^{U\times B}} \|\bI_U-\tilde\bW^H\bH\|_F^2 + \rho\|\tilde\bW^H\|^2_F
\end{align}
with regularization parameter $\rho=\No/\Es$. 
This optimization problem has a well-known closed-form solution given by \cite{paulraj03}
\begin{align} \label{eq:LMMSEmatrix}
\bW^H = (\rho\bI_U+\bH^H\bH)^{-1}\bH^H,
\end{align}
which can be computed efficiently in hardware \cite{studer2011asic}. 

We can alternatively compute the rows~$\bmw_u^H $, $u=1,\ldots,U$, of the L-MMSE equalization matrix  $\bW^H$ by decomposing \fref{eq:LMMSEmatrixProblem} into $U$ independent per-UE problems as follows:
\begin{align} \label{eq:MMSEequalizer}
\bmw_u  = \argmin_{\tilde\bmw\in\complexset^{B}} \|\bme_u-\bH^H\tilde\bmw\|_2^2 + \rho\|\tilde\bmw\|^2_2.
\end{align}
This alternative formulation of the L-MMSE equalizer will turn out useful in the next section.
\section{FAME: Finite-Alphabet MMSE Equalization}
\label{sec:FAME}
We now propose the finite-alphabet equalization paradigm.
We start by defining a finite-alphabet equalization matrix that enables efficient hardware for low-cardinality alphabets. 
We then formulate the FAME problem, which computes the finite-alphabet equalization matrix that minimizes the post-equalization MSE.
Finally, we present a simple approach to compute finite-alphabet equalization matrices and compare its performance to the one of an equalizer that solves the FAME problem exactly.
\subsection{Finite-Alphabet Equalization}
Linear equalization in hardware requires the computation of an inner product $\hat{s}_u = \langle \bmw_u,\bmy\rangle=\bmw_u^H\bmy$ per UE for every received vector~$\bmy$.
As demonstrated in \fref{sec:intro_example}, executing even such simple computations at the bandwidth offered by mmWave systems can result in excessively large area and high power consumption. 
To reduce both the area and power consumption, we propose to reduce the numerical precision of the equalization vectors $\bmw_u$, $u=1,\ldots,U$.
In the extreme case where the entries of $\bmw_u$ are quantized using 1-bit per real and imaginary component, an inner-product computation would only require additions and subtractions; this is  significantly less costly (in area and power) than using high-precision multipliers~\cite{Zimmerman99}.
However, it is obvious that reducing the numerical precision of the equalization vectors $\bmw_u$ will affect the MSE and eventually the error-rate performance.
Furthermore, quantization to, e.g., the finite alphabet $\setX=\{+1+j,+1-j,-1+j,-1-j\}$,  will result in numerical-range issues, meaning that such matrices will not be able to represent large or small entries.
To mitigate both of these issues, we now develop a principled way to perform equalization with what we call \emph{finite-alphabet matrices}.
Concretely, we are interested in designing equalization matrices with the structure defined next.
\begin{defi}\label{def:finitealphabetequalizationmatrix}
We define a $U\times B$ \emph{finite-alphabet equalization matrix}~as follows:
\begin{align} \label{eq:FAMEmatrix}
\bV^H = \mathrm{diag}(\boldsymbol\beta^*) \bX^H.
\end{align}
Here,  $\boldsymbol\beta\in\complexset^U$ is a vector that consists of post-equalization scaling factors and $\bX^H\in\setX^{U\times B}$ is an equalization matrix with entries chosen from the finite alphabet $\setX$. 
\end{defi}

\begin{rem}\label{rem:1bitfinitealphabet}
In this work, we are particularly interested in finite alphabets $\setX$ of low cardinality and whose elements can be represented using a small number of bits. An example  is the ``1-bit'' finite alphabet $\setX=\{+1+j,+1-j,-1+j,-1-j\}$, which uses $1$-bit per real and imaginary component.
\end{rem}

With \fref{def:finitealphabetequalizationmatrix}, the equalized received symbol for the $u$th UE is given by
\begin{align} \label{eq:equalization}
\hat s_u = \bmv^H_u\bmy =  \beta_u^* \bmx^H_u \bmy,
\end{align}
where  $\bmv^H_u\in\complexset^{1\times B}$ and $\bmx^H_u\in \setX^{1 \times B}$ are the $u$th rows of the matrices $\bV^H$ and $\bX^H$, respectively. 
It is now key to realize that spatial equalization as in \fref{eq:equalization} allows for efficient circuit implementations, especially for finite alphabets with low cardinality and regularly spaced elements. 
For such matrices, the inner product $\bmx^H_u \bmy$ can be implemented using low-resolution multipliers.
As $\boldsymbol\beta\in\complexset^U$, the post-equalization scaling operation by the scalar factor $\beta_u^*$ is performed using high-resolution multipliers.
Nonetheless, this operation is executed only once per UE.
In \fref{sec:vlsi}, we show that equalizer implementations that leverage finite-alphabet equalization matrices enable significant area and power savings.

\subsection{FAME: Finite-Alphabet MMSE Equalization}
We now propose FAME, a principled method to compute MSE-optimal finite-alphabet equalization matrices.
Analogous to the derivation of the \mbox{L-MMSE} equalizer in~\fref{sec:lmmse}, we are interested in minimizing the post-equalization MSE
\begin{align}
\label{eq:famse}
\textit{FA-MSE} = \Ex{\bms,\bmn}{\|\bV^H\bmy-\bms\|^2_2},
\end{align}
which differs from the MSE in \fref{eq:MSE} as now $\bV^H= \mathrm{diag}(\boldsymbol\beta^*) \bX^H$ is a finite-alphabet equalization matrix as per \fref{def:finitealphabetequalizationmatrix}.
The FAME matrix is the finite-alphabet equalization matrix that minimizes \fref{eq:famse}, i.e., it is the solution to the problem
\begin{align}
\label{eq:FAMSEmatrixProblem}
\bV^H  = \argmin_{\tilde\bX\in\setX^{U\times B},\tilde{\boldsymbol\beta}\in\complexset^U} \|\bI_U-\tilde\bV^H\bH\|_F^2 + \rho\|\tilde\bV^H\|^2_F,
\end{align}
where $\tilde\bV^H=\mathrm{diag}(\tilde{\boldsymbol\beta}^*) \tilde\bX^H$.
Clearly, the problem in \fref{eq:FAMSEmatrixProblem} mirrors the one in \fref{eq:LMMSEmatrixProblem}.
Hence, it follows from \fref{eq:MMSEequalizer} that the rows $\bmv^H_u=\beta^*_u\bmx_u^H$, $u=1,\ldots,U$, of the desired FAME matrix can be computed by solving the following optimization problem:
\begin{align} \label{eq:FAME}
\{\beta_u,\bmx_u\} = \argmin_{\tilde\bmx\in\setX^B,\tilde\beta\in\complexset}\|\bme_u- \bH^H\tilde\beta\tilde\bmx\|_2^2 + \rho \|\tilde\beta\tilde\bmx\|_2^2.
\end{align}
Intuitively, we are interested in finding the finite-alphabet equalization vectors $\bmv^H_u=\beta^*_u\bmx^H_u$, $u=1,\ldots,U$, that best mimic the infinite-precision L-MMSE equalizer. 
\begin{rem}
For a fixed scaling factor $\beta_u$, the FAME problem in~\fref{eq:FAME} is a closest vector problem, which is known to be NP-hard \cite{agrell02a,fincke85a,verdu89a}. 
For example, for a system with $256$ BS antennas using a 1-bit finite-alphabet equalization matrix, solving the FAME problem using an exhaustive search would require one to evaluate the objective function in \fref{eq:FAME} more than $10^{154}$ times for each UE.
Clearly, without low-complexity algorithms, the FAME problem cannot be solved in practical massive  MU-MIMO  mmWave systems.
\end{rem}

Since the FAME problem in \fref{eq:FAME} minimizes the cost function for two quantities at once, i.e., the scaling factor $\beta_u$ and the low-resolution vector $\bmx_u$, it is not obvious how to solve it efficiently. To derive computationally efficient algorithms in \fref{sec:lowBitFAME}, we will use the following equivalent form of the FAME problem in \fref{eq:FAME}; the proof is given in \fref{app:FAMEequivalent}.
\begin{lem} \label{lem:FAMEequivalent}
The FAME problem in \fref{eq:FAME} is equivalent to solving the following optimization problem for each UE $u=1,\ldots,U$:
\begin{align} \label{eq:FAMEcompact}
\bmx_u = \argmin_{\tilde\bmx\in\setX^B} \frac{\|\bH^H\tilde\bmx\|_2^2+\rho\|\tilde\bmx\|_2^2}{|\bmh^H_u\tilde\bmx|^2 },
\end{align}
where the associated optimal scaling factor is given by
\begin{align} \label{eq:optimalscaling}
\beta_u(\vecx_u)
= \frac{ \bmx^H_u\bmh_u }{\|\bH^H\bmx_u\|_2^2 + \rho\|\bmx_u\|_2^2}.
\end{align}
\end{lem}
This formulation of the FAME problem allows us  to first find the optimal vector $\bmx_u$ using \fref{eq:FAMEcompact} and then compute the associated optimal scaling factor $\beta_u$ using \fref{eq:optimalscaling}.
Note that the equation in \fref{eq:optimalscaling} represents the MSE-optimal scaling factor $\beta_u$ for a given vector $\bmx_u$, regardless of how $\bmx_u$ was computed.
Furthermore, the equivalent formulation in \fref{lem:FAMEequivalent} is similar to a formulation proposed in~\cite{jacobsson17d} in the context of nonlinear quantized precoders.

\subsection{FL-MMSE: A Baseline Finite-Alphabet Equalizer}
\label{sec:qlmmse}
Since the FAME problem is NP-hard, we now present a baseline method to compute finite-alphabet equalization matrices as per~\fref{def:finitealphabetequalizationmatrix} without having to solve the FAME problem in~\fref{eq:FAMEcompact}.
We call our approach finite-alphabet \mbox{L-MMSE} (FL-MMSE), as it obtains the entries of the low-resolution matrix~$\bX^H$ by quantizing the L-MMSE equalizer in \fref{eq:LMMSEmatrix}---the corresponding scaling factors $\beta_u$, $u=1,\ldots,U$, are then obtained via \fref{eq:optimalscaling}.
In this work, we will use the FL-MMSE equalizer as a baseline to evaluate the performance of other, more sophisticated finite-alphabet equalizers that attempt to directly solve the FAME problem~in~\fref{eq:FAMEcompact}.

For the $1$-bit case, FL-MMSE applies the signum function~$\text{sgn}(\cdot)$ separately on the real and imaginary parts of the L-MMSE matrix $\bW^H$ to obtain $\bX^H$.
Then, FL-MMSE uses the expression in \fref{eq:optimalscaling} to compute the high-resolution scaling factors in the vector $\boldsymbol\beta$.
FL-MMSE can also be used with finite alphabets that have more than $1$-bit per complex entry.
In such cases, after computing the L-MMSE equalization matrix~$\bW^H$ in~\fref{eq:LMMSEmatrix}, we proceed to quantize its real and imaginary parts as follows.
First, for each row $\bmw_u^H$ of $\bW^H$, we identify the scalar $w_{\text{max}}$ corresponding to the largest absolute value in $[\Re\{\bmw_u^H\},\Im\{\bmw_u^H\}]$.
Then, assuming that the targeted resolution is $r$ bits, we divide the range  $[-w_{\text{max}},+w_{\text{max}}]$ into $2^r$ uniform-width bins and quantize the entries of $\Re\{\bmw_u^H\}$ and $\Im\{\bmw_u^H\}$ to the centroid values of these bins.
For $2$-bit resolution, for example, the centroid values of the bins are $\{-0.75,-0.25,+0.25,+0.75\}w_{\text{max}}$.
In hardware, one would scale these centroid values so that the minimum absolute value corresponds to $1$. Following the previous example, one would use the values $\{-3,-1,+1,+3\}$ to represent the entries of $\Re\{\bmx_u^H\}$ and $\Im\{\bmx_u^H\}$.
Note that this scaling does not affect the solution of the FAME problem in \fref{eq:FAMEcompact}, as it is absorbed by the scaling factor $\beta_u$ in \fref{eq:optimalscaling}.
After obtaining the low-resolution vector $\bmx_u^H$, the corresponding scaling factor $\beta_u$ is computed using the expression in \fref{eq:optimalscaling}.

\begin{figure*}[tp]
\centering
\subfigure[L-MMSE ({$\textit{EVM}=11.58\%$})]{\includegraphics[width=.275\textwidth]{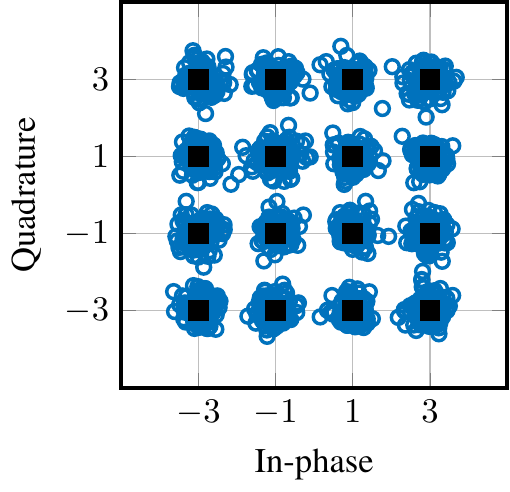}\label{fig:scatter_lmmse}}
\hfill
\subfigure[FL-MMSE ({$\textit{EVM}=30.58\%$})]{\includegraphics[width=.275\textwidth]{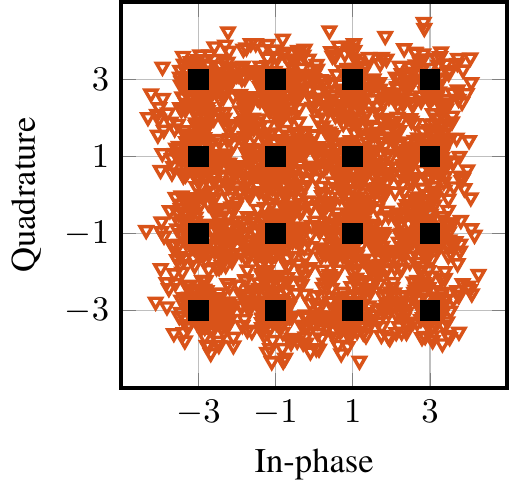}\label{fig:scatter_qlmmse}}
\hfill
\subfigure[FAME-EXH ({$\textit{EVM}=15.30\%$})]{\includegraphics[width=.275\textwidth]{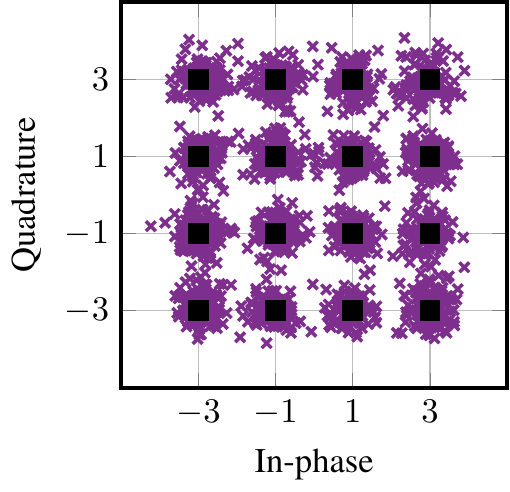}\label{fig:scatter_fame}}
\caption{Error-vector magnitude (EVM) performance for different equalizers in a $B=8$ BS antenna, $U=2$ UE, $16$-QAM system with i.i.d.\ Rayleigh-fading channels at $15$\,dB SNR. FL-MMSE corresponds to 1-bit quantization of the L-MMSE equalizer, which results in a significant EVM degradation. Solving the 1-bit  FAME problem using an exhaustive search (FAME-EXH) yields an EVM comparable to that of infinite-precision L-MMSE equalization.}\label{fig:scatter_all}
\end{figure*}

\begin{figure*}[tp]
\centering
\subfigure[L-MMSE ({$\textit{SINR}=21.02$\,dB})]{\includegraphics[width=.3\textwidth]{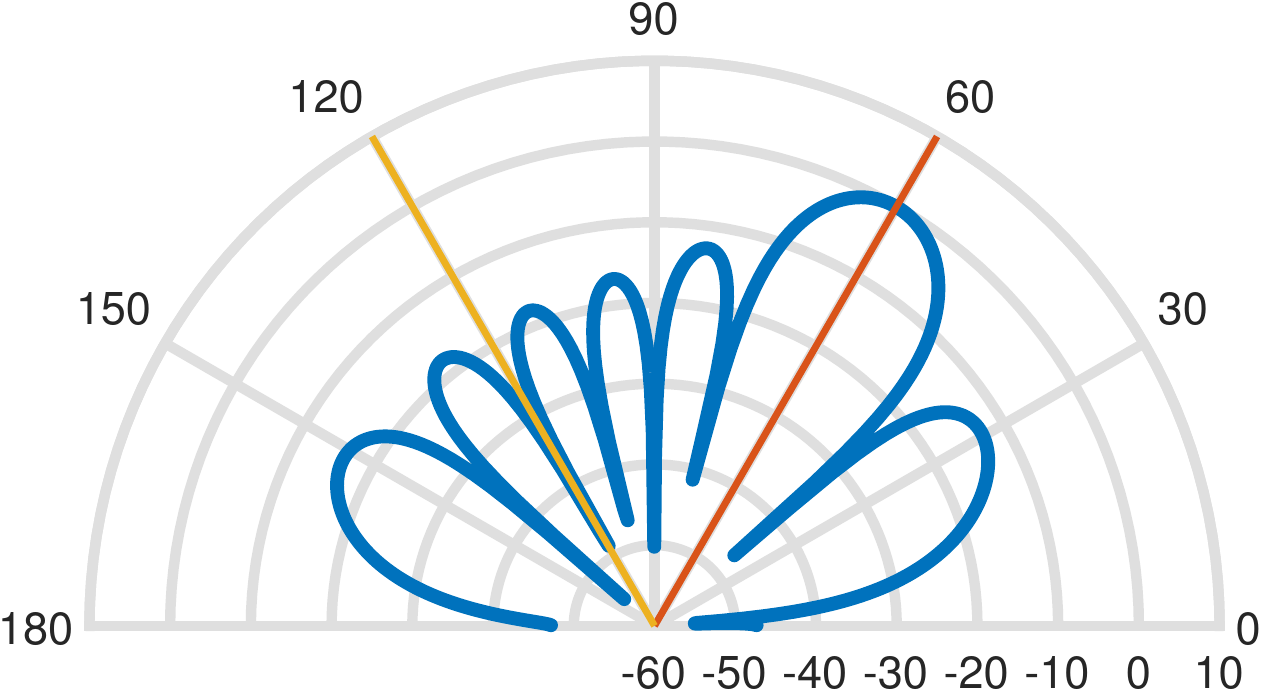}\label{fig:angular_lmmse}}
\hfill
\subfigure[FL-MMSE ({$\textit{SINR}=6.72$\,dB})]{\includegraphics[width=.3\textwidth]{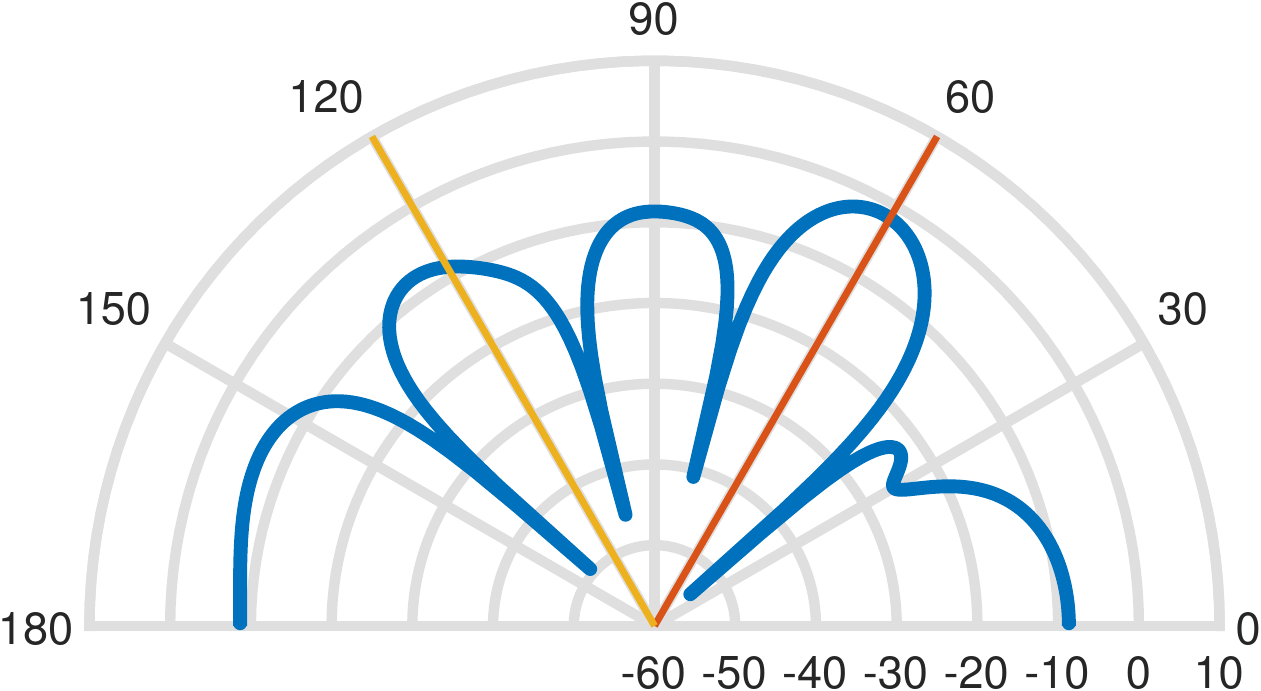}
\label{fig:angular_qlmmse}}
\hfill
\subfigure[FAME-EXH ({$\textit{SINR}=21.02$\,dB})]{\includegraphics[width=.3\textwidth]{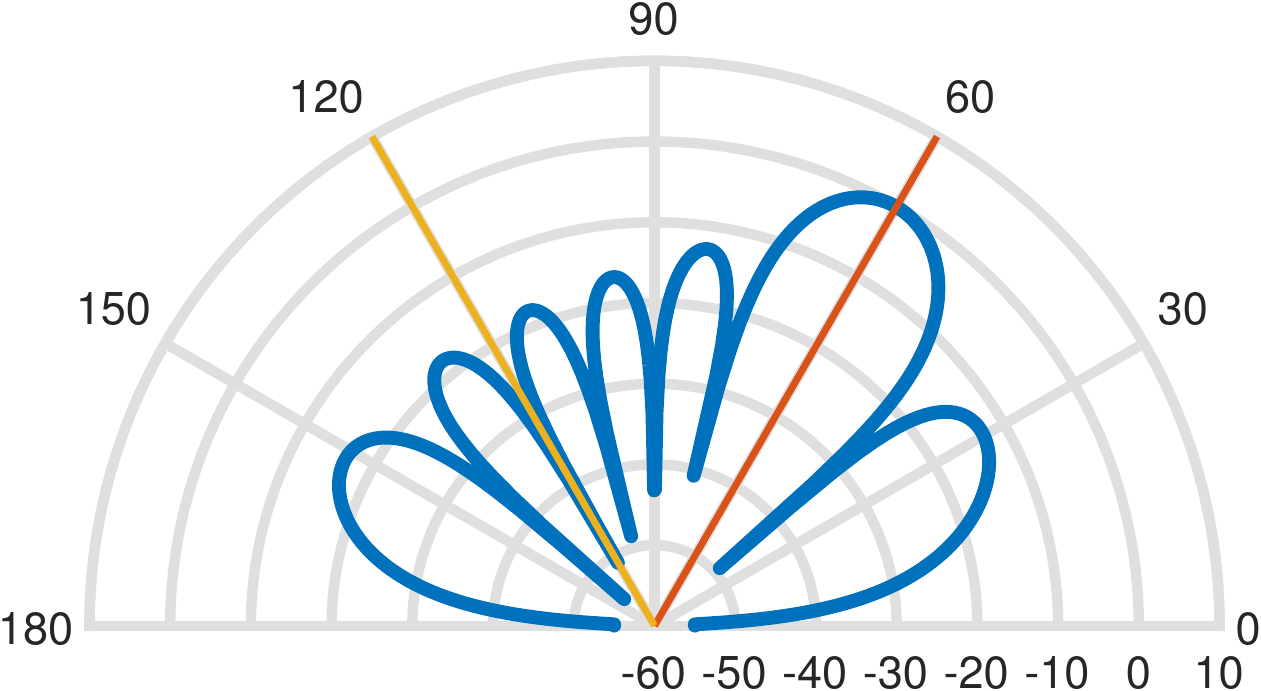}\label{fig:angular_fame}}
\caption{Beam- and null-forming capabilities for different equalizers in a $B=8$ BS antenna, $U=2$ UE system for textbook LoS channel at an SNR of $15$\,dB. The primary UE is located at $60^\circ$ with respect to the ULA of antennas, whereas the secondary UE is at $120^\circ$. We show the received signal power in decibels of the signals coming from different angles after being equalized for the primary UE at $60^\circ$. A good equalizer gathers energy from this UE, while canceling interference from the UE at $120^\circ$. FL-MMSE, which performs $1$-bit quantization on the L-MMSE matrix, fails to reject the interference from the $120^\circ$ UE. In contrast, $1$-bit FAME-EXH achieves nearly identical beam- and null-forming performance to that of infinite-precision L-MMSE equalization.}
\label{fig:angular_all}
\end{figure*}

\subsection{EVM and Beamforming Performance of Exact FAME}
\label{sec:fame_exh_exp}
We now assess the EVM and beamforming performance of the optimal FAME problem in~\fref{eq:FAMEcompact} for a $1$-bit finite alphabet. To solve the NP-hard problem, we resort to an exhaustive search, which we call FAME-EXH.
To keep the complexity within reasonable bounds, we simulate a small MU-MIMO system with $B=8$ BS antennas.
We also simulate the performance of conventional, infinite-precision L-MMSE equalization and 1-bit  FL-MMSE equalization as proposed in \fref{sec:qlmmse}.

\subsubsection{Error-Vector Magnitude}
\fref{fig:scatter_all} shows scatter plots of the equalization outputs $\hat{\bms}$ of L-MMSE, FL-MMSE, and FAME-EXH equalization for 2\,000 realizations in a $B=8$ BS antenna, $U=2$ UE system operating over an i.i.d. Rayleigh-fading channel at $15$\,dB SNR.
While the infinite-precision L-MMSE equalizer achieves an EVM of {$11.58$\%}, quantizing its solution to 1-bit using FL-MMSE degrades the EVM to {$30.58$\%}, which blurs the decision regions of the considered 16-QAM constellation.
In stark contrast, the $1$-bit FAME-EXH equalizer achieves an EVM of only {$15.30$\%}, which is close to that of the infinite-precision L-MMSE equalizer; furthermore, the decision regions between constellation points are clearly visible. 
These results demonstrate the significant EVM advantage of solving the FAME problem over the simple FL-MMSE equalizer. 

\subsubsection{Beam- and Null-forming Performance}
\fref{fig:angular_all} illustrates the beam- and null-forming capabilities of FAME-EXH.
For these plots, we consider a $B=8$ BS antenna, $U=2$ UE system operating at  $15$\,dB  SNR over a textbook LoS channel, where the channel coefficient between the $b$th BS antenna and an UE located at an angle of $\phi$ is modeled as follows \cite{tse05a}:
\begin{equation}
\label{eq:vlos_model}
h_b(\phi) = e^{-j\pi(b-1)\cos(\phi)}, \quad b=1,\ldots, B.
\end{equation}
Here, we assume a uniform linear array (ULA) of antennas with half-wavelength antenna spacing and constant path loss.
We locate a primary UE at an angle of $\phi_1=60^\circ$ and a secondary UE at $\phi_2=120^\circ$. Next, we compute the corresponding equalization matrix using L-MMSE, FL-MMSE, and FAME-EXH equalization.
We then evaluate how much the equalization vector $\bmv_1^H$ (which corresponds to the UE at $\phi_1=60^\circ$) captures (or rejects) signals incoming from different incident angles by evaluating $|\bmv_1^H \bmh(\phi')|^2$ for $0\le\phi'\le\pi$.
The equalization vector~$\bmv_1^H$ should amplify the signal from the primary UE at $\phi_1=60^\circ$ but attenuate the signal from the secondary UE.
The results shown in \fref{fig:angular_all} demonstrate that infinite-precision L-MMSE equalization is able to simultaneously beam-form towards the primary UE and null interference from the secondary UE. The $1$-bit FL-MMSE equalizer is unable to reject interference from the secondary UE. 
In stark contrast, $1$-bit FAME-EXH equalization is able to both beam-form towards the primary UE and null-form towards the secondary UE. 
We confirm these observations by computing the signal-to-interference-plus-noise ratio (SINR) for the primary UE.

Despite the significant performance advantages of $1$-bit FAME-EXH over $1$-bit FL-MMSE, solving $1$-bit FAME-EXH for large-dimensional problems that arise in mmWave systems is infeasible in practice. To this end, we next develop low-complexity FAME solvers that scale to large BS antenna arrays.
\section{Fast Algorithms to Solve FAME}
\label{sec:lowBitFAME}
We next present approximate algorithms to solve the FAME problem efficiently for systems with a large number of BS antennas.
We start by proposing a semidefinite relaxation (SDR)-based method and then develop a much faster method that uses forward-backward splitting (FBS). 

\subsection{FAME with Semidefinite Relaxation (SDR)}
We focus on using SDR~\cite{luo10sdr} to solve the FAME problem in~\fref{eq:FAMEcompact} for a $1$-bit finite alphabet.
To do so, we first re-express the FAME problem in the real domain using the quantities
\begin{align}
\bmx_\reals = \left[
\begin{array}{c}
\Re\{\bmx\} \\
\Im\{\bmx\} 
\end{array}
\right]
\,\,\, \text{and}\,\,\,\,
\bH_\reals = \left[
\begin{array}{cc}
\Re\{\bH\} & -\Im\{\bH\} \\
\Im\{\bH\} & \Re\{\bH\} 
\end{array}
\right]\!.
\end{align}
Throughout, we will assume that $\Re\{\bmx\}$ and $\Im\{\bmx\}$ take values from the same alphabet $\setX_\reals$. 
For example, for 1-bit finite alphabets, we have $\setX_\reals=\{-1,+1\}$.
We can now rewrite the FAME problem in \fref{eq:FAMEcompact} as 
\begin{align} \label{eq:FAMEcompactreal}
\bmx_{\reals,u} = \argmin_{\tilde\bmx_\reals\in\setX^{2B}_\reals} \frac{\|\bH^H_\reals\tilde\bmx_\reals\|_2^2+\rho\|\tilde\bmx_\reals\|_2^2}{|\bmh^H_{\reals,u}\tilde\bmx_\reals|^2 }.
\end{align}
It is now key to realize that the vector $\tilde\bmx_\reals$ can be scaled arbitrarily  without changing the objective function of \fref{eq:FAMEcompactreal}.
This observation enables us to state an equivalent optimization problem
\begin{align}
\label{eq:normtrickSDR}
\left\{\begin{array}{ll}
\underset{\bar\bmx_\reals\in\setZ^{2B}_\alpha,\alpha>0}{\text{minimize}} &   \|\bH^H_\reals\bar\bmx_\reals\|_2^2+\rho\|\bar\bmx_\reals\|_2^2  \\ 
\text{subject to } & |\bmh^H_{\reals,u}\bar\bmx_\reals|^2=1,
\end{array}
\right.
\end{align}
where the discrete set $\setZ_\alpha$ is a scaled version of $\setX_\reals$; for 1-bit finite alphabets, we have $\setZ_\alpha=\{-\alpha,+\alpha\}$ with $\alpha>0$. This formulation enables us to formulate a semidefinite program to solve the FAME problem approximately. 

By focusing on 1-bit finite alphabets, we can relax \fref{eq:normtrickSDR} by replacing the constraint $|\bmh^H_{\reals,u}\bar\bmx_\reals|^2=1$ by \mbox{$ \bmh^H_{\reals,u}\bar\bX \bmh_{\reals,u} = 1$}, where the positive semidefinite matrix $\bar\bX\in\opS^{2B}_+$ should approximate  $\bar\bmx_\reals\bar\bmx_\reals^H$. This SDR yields 
\begin{align}
\label{eq:FAMESDRproblem}
\left\{\begin{array}{ll}
\underset{\bar\bX\in\opS^{2B}_+}{\text{minimize}} &  \tr\big((\bH_\reals\bH_\reals^H+\rho\bI_{2B})\bar\bX\big)   \\ 
\text{subject to } & \bmh^H_{\reals,u}\bar\bX \bmh_{\reals,u} = 1\\
& \bar X_{1,1}= \bar X_{b,b}, b=2,\ldots,2B,
\end{array}
\right.
\end{align}
where we ensure that the diagonal elements of $\bar\bX\in\setS^{2B}_+$ must be equal (but we do not specify their value). This constraint is a result of the fact that we are interested in a solution in the set~$\setZ_\alpha$, where the parameter $\alpha$ is not known. 
After solving the semidefinite program in \fref{eq:FAMESDRproblem}, we compute the finite-alphabet vector by first extracting the leading eigenvector of the solution matrix $\bar\bX$ followed by quantizing it to $\{-1,+1\}$ using the signum function~$\text{sgn}(\cdot)$. The equalization vector can then be scaled using the optimal FAME  scaling parameter in \fref{eq:optimalscaling}. 
We refer to this procedure as FAME-SDR. A more detailed description of general SDR techniques can be found in \cite{luo10sdr}.

\begin{rem}
While FAME-SDR can also be derived for  multi-bit finite alphabets, e.g., using the techniques described in~\cite{chang2009linear}, we will not pursue this approach for the following reasons.
As we will show in \fref{sec:compl}, the complexity of FAME-SDR does not scale well to a large number of BS antennas.
Moreover, FAME-SDR cannot be applied to finite alphabets that are not separable into  real and imaginary parts, such as a finite alphabet that contains the  elements of an 8-PSK (phase shift keying) constellation.
In addition, to the best of our knowledge, SDR can only handle finite alphabets with even cardinality that exclude a zero element. 
To avoid the drawbacks of SDR for FAME, we next present an alternative approach.
\end{rem}

\subsection{FAME with Forward-Backward Splitting (FBS)}
\label{sec:FAMEFBS}
Due to the high complexity of FAME-SDR and the fact that SDR solvers are notoriously difficult to implement in hardware~\cite{castaneda16a}, we next develop a low-complexity alternative for solving the FAME problem approximately.
To do so, we start by assuming that, for each UE $u=1,\dots,U$, we know the optimal value of the objective in \fref{eq:FAMEcompact}, which we denote with $\gamma_u$. Mathematically,
\begin{equation}
\label{eq:gammaDef}
\gamma_u=\frac{\|\bH^H\bmx_u\|_2^2+\rho\|\bmx_u\|_2^2}{|\bmh^H_u\bmx_u|^2 },
\end{equation}
where $\bmx_u$ is the solution to the problem in \fref{eq:FAMEcompact}.
Note that it follows from~\fref{eq:gammaDef} that~$\gamma_u>1$.
By rearranging~\fref{eq:gammaDef}, we obtain
\begin{equation}
0=\|\bH^H\bmx_u\|_2^2+\rho\|\bmx_u\|_2^2-\gamma_u|\bmh^H_u\bmx_u|^2.
\end{equation}
Thus, if $\gamma_u$ was known, solving the problem 
\begin{equation}
\label{eq:FAME2solve}
\bmx_u=\argmin_{\tilde{\bmx}\in\setX^B}\frac{1}{2}\|\bH^H\tilde\bmx\|_2^2+\frac{\rho}{2}\|\tilde\bmx\|_2^2-\frac{\gamma_u}{2}|\bmh^H_u\tilde\bmx|^2
\end{equation}
would yield the same solution as \fref{eq:FAMEcompact}.
As the optimal value of the objective in~\fref{eq:FAMEcompact} is unknown in practice, we will use $\gamma_u$ as an algorithm parameter that we tune to empirically improve the algorithm's performance.

Since the problem in \fref{eq:FAME2solve} still contains a search over the finite alphabet $\setX^B$, we relax the non-convex constraint $\tilde{\bmx}\in\setX^B$ to $\tilde{\bmx}\in\setB^B$.
Here, $\setB$ corresponds to the convex hull of the finite alphabet~$\setX$, which is defined as \cite{wu2016high}
\begin{equation}
\setB = \left\lbrace
 \sum_{i=1}^{|\setX|} \alpha_i \bar x_i \mid (\alpha_i \in \reals_+, \forall i )
 \wedge
 \sum_{i=1}^{|\setX|} \alpha_i = 1
\right\rbrace\!,
\end{equation}
where $\bar x_i$ is the $i$th element of $\setX$ and $i=1,\ldots,|\setX|$.
After this relaxation step, the all-zeros vector $\bm0_{B\times1}$ becomes a trivial solution.
To prevent the algorithm from returning this trivial  solution, we follow the approach put forward in \cite{shah2016biconvex} and include a term in \fref{eq:FAME2solve} that encourages large entries in the vector~$\tilde\bmx$.
Specifically,  we add $-\frac{\delta}{2}\|\tilde\bmx\|_2^2$ to the objective function, where $\delta>0$ is a regularization parameter. 
The resulting optimization problem is given by
\begin{align}
\label{eq:FAME_BCR}
\bmx_u=\argmin_{\tilde{\bmx}\in\setB^B}\frac{1}{2}\|\bH^H\tilde\bmx\|_2^2-\frac{\gamma_u}{2}|\bmh^H_u\tilde\bmx|^2+\frac{\rho-\delta}{2}\|\tilde\bmx\|_2^2.
\end{align}

To compute a solution to \fref{eq:FAME_BCR}, we utilize FBS~\cite{BT09,GS10,GSB14}.
FBS is an efficient, iterative solver for convex optimization problems of the form
\begin{align}
\label{eq:FBSform}
\hat{\bmx} = \argmin_{\tilde\bmx} f(\tilde\bmx) + g(\tilde\bmx),
\end{align}
where both functions $f$ and $g$ are convex, but $f$ is a smooth function and $g$ is not necessarily smooth or bounded.  
FBS executes the following operations for $t=1,2,\dots,t_\text{max}$ iterations or until convergence \cite{BT09,GSB14}:
\begin{align}
\tilde\bmz^{(t+1)} & =\tilde\bmx^{(t)}-\tau^{(t)}\nabla f(\tilde\bmx^{(t)}) \label{eq:fbs_z}\\
\tilde\bmx^{(t+1)} & =\text{prox}_g\left(\tilde\bmz^{(t+1)};\tau^{(t)}\right).
\end{align}
Here, $\nabla f(\tilde\bmx^{(t)})$ is the gradient of the function $f$, $\lbrace\tau^{(t)}>0\rbrace$ is a sequence of step sizes, and $\text{prox}_g(\cdot)$ is the proximal operator of the function $g$, defined as \cite{parikh2014proximal}
\begin{align}
\text{prox}_g\left(\tilde\bmz;\tau\right) = \argmin_{\tilde\bmx} \left\lbrace \tau g(\tilde\bmx) + \frac{1}{2}\|\tilde\bmx-\tilde\bmz\|^2\right\rbrace.
\end{align}

Our problem in \fref{eq:FAME_BCR} is not convex and hence, FBS is not guaranteed to converge to an optimal solution. Nevertheless, we can use FBS to find approximate solutions to  \fref{eq:FAME_BCR} by setting 
\begin{align}
f(\tilde\bmx) & = \frac{1}{2}\|\bH^H\tilde\bmx\|_2^2-\frac{\gamma_u}{2}|\bmh^H_u\tilde\bmx|^2\\
g(\tilde\bmx) & = \bm1_{\setB^B}(\tilde\bmx)+\frac{\rho-\delta}{2}\|\tilde\bmx\|^2_2.
\end{align}
Here, the convex constraint $\tilde{\bmx}\in\setB^B$ in \fref{eq:FAME_BCR} is incorporated into the function $g(\tilde\bmx)$ via the indicator function $\bm1_{\setB^B}(\tilde\bmx)$, which is zero if $\tilde\bmx\in\setB^B$ and infinity otherwise.
With these definitions, we arrive at
\begin{align}
\nabla f(\tilde\bmx) = &~\bH\bH^H\tilde\bmx-\gamma_u\bmh_u\bmh^H_u\tilde\bmx \label{eq:fbs_grad}\\
\text{prox}_g\left(\tilde{z}\right) = &~\text{sgn}\left(\Re\lbrace\tilde{z}\rbrace\right)\min\left\{\nu^{(t)}|\Re\lbrace\tilde{z}\rbrace|,1\right\} \nonumber \\
& + j~\text{sgn}\left(\Im\lbrace\tilde{z}\rbrace\right)\min\left\{\nu^{(t)}|\Im\lbrace\tilde{z}\rbrace|,1\right\} \label{eq:proxgFAME},
\end{align}
where $\nu^{(t)}=(1+\tau^{(t)}(\rho-\delta))^{-1}$ and \fref{eq:proxgFAME} is applied element-wise to the vector $\tilde\bmz$.

Note that we have introduced three sets of algorithm parameters: $\{\tau^{(t)}\}$, $\{\nu^{(t)}\}$, and $\{\gamma_u\}$, where $t=1,\ldots,t_{\text{max}}$ and $u=1,\ldots,U$.
While one could manually tune these parameters via numerical simulations, we automate the tuning process by using a neural-network-based approach as put forward in \cite{balatsoukas-stimming19a}.
We note that such a neural-network-based approach is only used for determining the algorithm parameters, i.e., it is trained offline without affecting the runtime complexity of FAME-FBS.
As the same algorithm parameters should work across several channel realizations, having a per-UE parameter such as $\{\gamma_u\}$ is meaningless.
As a result, we set $\gamma=\gamma_u$ for $u=1,\ldots,U$.
Furthermore, to provide the neural network with greater flexibility during optimization, we allow~$\gamma$ to be different in each iteration; i.e., we introduce another set of per-iteration parameters $\{\gamma^{(t)}\}$, $t=1,\ldots,t_\text{max}$.
We call the resulting algorithm FAME-FBS, which is summarized~next.

\begin{oframed}
\vspace{-0.25cm}
\begin{alg}[FAME-FBS]\label{alg:FAME-FBS} 
Initialize \mbox{$\tilde\bmx^{(1)}$} with either the maximum-ratio combining (MRC) solution \mbox{$\bmh_u$} or the low-resolution vector \mbox{$\bmx_u$} computed by FL-MMSE, and fix the sets of parameters {$\{\tau^{(t)}\}$}, {$\{\nu^{(t)}\}$}, and {$\{\gamma^{(t)}\}$}.
Then, for every iteration $t=1,2,\ldots,t_\text{max},$ compute 
\begin{align}
\tilde\bmz^{(t+1)} &= \left(\bI_B - \tau^{(t)}\bH(\bI_U-\gamma^{(t)}\bme_u\bme_u^H)\bH^H\right) \tilde\bmx^{(t)} \label{eq:step1}\\
\tilde\bmx^{(t+1)} & = \mathrm{prox}_g(\tilde\bmz^{(t+1)}). \label{eq:step2}
\end{align}
Here, the proximal operator $\mathrm{prox}_g(\cdot)$ is the element-wise function given by \fref{eq:proxgFAME}.
The result $\tilde\bmx^{(t_\text{max}+1)}$ is quantized to the finite alphabet $\setX$ to obtain $\bmx_u^H$.
Then, the optimal FAME scaling parameter $\beta_u$ is computed using \fref{eq:optimalscaling}.
\end{alg}
\vspace{-0.25cm}
\end{oframed}

\begin{rem}
FAME-FBS supports multi-bit finite-alphabet equalization matrices.
This is achieved by uniformly quantizing, in the range $[-1,+1]$, the real and imaginary parts of the solution vector $\tilde\bmx^{(t_\text{max}+1)}$, similar to what is done by FL-MMSE equalization.
As a consequence, unlike FAME-SDR, FAME-FBS can operate with finite alphabets that contain (i) an odd number of elements and (ii) a zero element.
Furthermore, FAME-FBS (and FL-MMSE) can be applied to PSK-like finite alphabets following an approach related to the one used~in~\cite{castaneda18c3po}.
\end{rem}

\begin{rem}
Since FAME-FBS was obtained by relaxing the original optimization problem and by applying an iterative solver for convex optimization problems to the non-convex problem in~\fref{eq:FAME_BCR}, it is not guaranteed to converge to the optimal solution of the original problem in \fref{eq:FAMEcompact}.
Nonetheless, our simulation results in \fref{sec:sim} confirm that FAME-FBS achieves competitive performance for (i) different system configurations and (ii) realistic channel models.
\end{rem}

\subsection{Computational Complexity}
\label{sec:compl}

We now assess the complexity of (i) computing the equalization matrix and (ii) performing equalization on a received vector $\bmy$, for high-resolution and finite-alphabet equalization approaches.
We measure computational complexity as the number of real-valued multiplications performed by an algorithm.\footnote{For the remainder of the paper, we assume that complex-valued multiplications require four real-valued multiplications.}

\begin{table}[tp]
\centering
{\caption{Complexity for computing an equalization matrix}
\label{tbl:complexitymatrix}
\renewcommand{\arraystretch}{1.1}
\begin{minipage}[c]{1\columnwidth}
\centering
\begin{tabular}{@{}llc@{}}
\toprule
{Algorithm} & Computational complexity & Asymptotic scaling \tabularnewline
\midrule
L-MMSE & $2U^3\!+\!6BU^2\!-\!2BU\!-\!2U\!+\!1$ & $O(BU^2)$  \tabularnewline
FL-MMSE & $10BU^2\!+\!2U^3\!+\!2U^2\!+\!U\!+\!1$ & $O(BU^2)$ \tabularnewline
FAME-SDR & n.a.  & $O(B^{4.5})$ \tabularnewline
\multirow{2}{*}{FAME-FBS} & $(8t_\text{max}\!+\!4)BU^2\!+\!2U^2$ & \multirow{2}{*}{$O(BU^2)$}  \tabularnewline
& $+(4t_\text{max}\!+\!2)BU\!+\!(2t_\text{max}\!+\!3)U$ & \tabularnewline
\bottomrule
\end{tabular}
\end{minipage}}
\end{table}

\subsubsection{Computing the Equalization Matrix}
\fref{tbl:complexitymatrix} lists the computational complexity for computing a single equalization matrix using L-MMSE, FL-MMSE, FAME-SDR, and FAME-FBS.
For the infinite-precision L-MMSE equalizer, the complexity corresponds to explicitly computing  the equalization matrix $\bW^H$.
For the finite-alphabet equalizers (FL-MMSE and FAME-based algorithms), the complexity corresponds to the computation of the low-resolution matrix $\bX^H$ and the scaling factors  in the vector~$\boldsymbol\beta$.
Solving FAME-SDR results in the highest complexity, which asymptotically scales as $O(B^{4.5})$ unless specific problem structures can be exploited~\cite{luo10sdr}. Since we do not have access to a particular SDR solver, we only provide this asymptotic scaling. 
Evidently, FAME-SDR does not scale well to systems with a large number of BS antennas. 
FAME-FBS has the same asymptotic scaling of $O(BU^2)$ as L-MMSE and FL-MMSE equalization, making it suitable for massive  MU-MIMO mmWave systems.
The exact complexity counts listed in \fref{tbl:complexitymatrix} are derived in \fref{app:count}.

\begin{rem}
While the constant associated with the term $BU^2$ is larger for FAME-FBS than for L-MMSE and FL-MMSE, the complexity of the latter algorithms appears to be higher in practice. 
Computing the L-MMSE (and the FL-MMSE) equalizer in hardware requires square roots and divisions, which result in high numerical precision requirements~\cite{studer2011asic}.
Furthermore, the Cholesky decomposition and forward- and back-substitution procedures required when computing the L-MMSE (and the FL-MMSE) equalization matrix result in stringent data dependencies that limit parallelism and, hence, reduce throughput. 
In contrast, FAME-FBS has a regular structure with few data dependencies and the matrix-vector multiplications can be parallelized easily. In addition, one can parallelize computation per UE as the FAME problem in \fref{eq:FAMEcompact} is independent for each $u=1,\ldots,U$. 
In fact, a simple hardware engine, similar to the one proposed in \cite{castaneda17a} for another FBS-based algorithm, could be used to efficiently execute FAME-FBS to determine the low-resolution equalization vectors $\bmx_u^H$. 
\end{rem}

\begin{table}[tp]
\centering
{\caption{Complexity of finite-alphabet equalization}
\label{tbl:complexityeq}
\renewcommand{\arraystretch}{1.1}
\begin{minipage}[c]{1\columnwidth}
\centering
\begin{tabular}{@{}lcc@{}}
\toprule
\multirow{2}{*}{Equalization} & \multicolumn{2}{c}{\,\,Real-valued multiplication count}\,\tabularnewline
& {High resolution} & {Low resolution} \tabularnewline
\midrule
Traditional & $4BU$ & $0$  \tabularnewline
Finite alphabet & $4U$ & $4BU$  \tabularnewline
\bottomrule
\end{tabular}
\end{minipage}}
\end{table} 

\subsubsection{Performing Equalization}
After computing the equalization matrix, one must perform spatial equalization on the received signal vectors $\bmy$ at the rate of the ADCs.
For the infinite-precision L-MMSE equalizer, this corresponds to computing one high-resolution matrix-vector product $\hat\bms=\bW^H\bmy$ per receive vector.
For finite-alphabet equalizers, this corresponds to a low-resolution matrix-vector product $\bmz=\bX^H\bmy$, followed by $U$ high-resolution products $\hat s_u =\beta_u^*z_u$, $u=1,\ldots,U$.
The complexity of equalization is summarized in \fref{tbl:complexityeq}, where we distinguish between high resolution and low resolution multiplications. 
While finite-alphabet equalization performs more multiplications than a conventional equalizer, most of these multiplications are performed at low resolution.
Thus, for sufficiently low resolution, finite-alphabet equalization effectively reduces the complexity of spatial equalization.

\begin{rem}
While spatial equalization must be carried out at symbol rate, the computation of the equalization matrix must only be carried out if the channel matrix changes.
Given that we are considering operation at extremely high bandwidths, the complexity of performing equalization will dominate in most mmWave systems. 
For scenarios with short coherence times, methods that minimize the complexity of computing the equalization matrix are to be preferred.
\end{rem}

\begin{figure}[tp]
\centering
\subfigure[$B=8$ BS antennas, $U=2$ UEs, QPSK]{\includegraphics[width=.85\columnwidth]{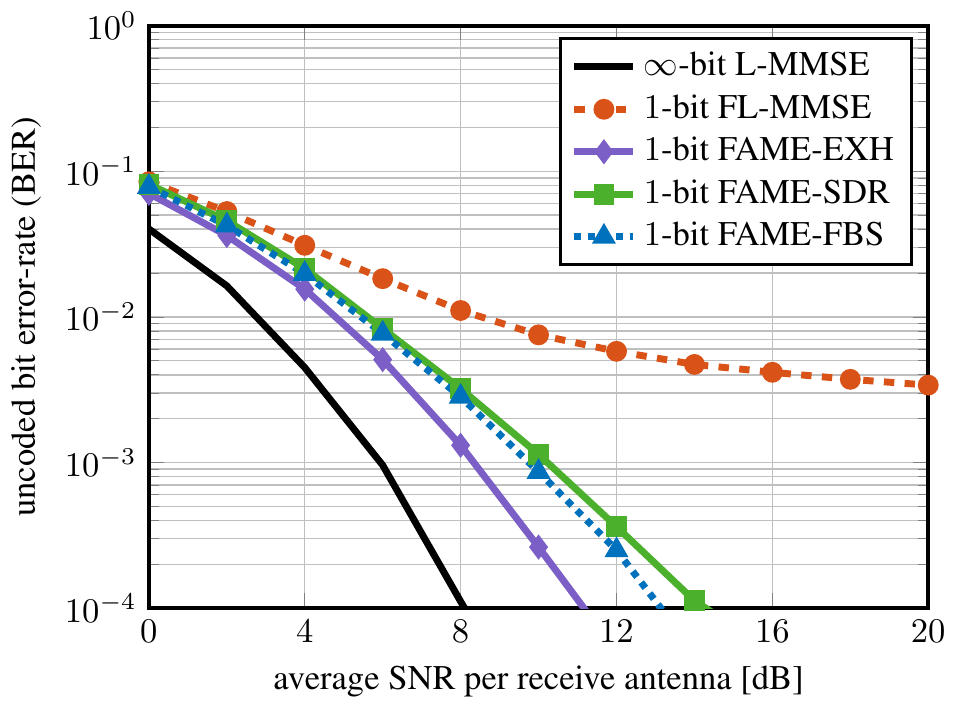}\label{fig:ber_8x2}} \\
\subfigure[$B=64$ BS antennas, $U=4$ UEs, $16$-QAM]{\includegraphics[width=.85\columnwidth]{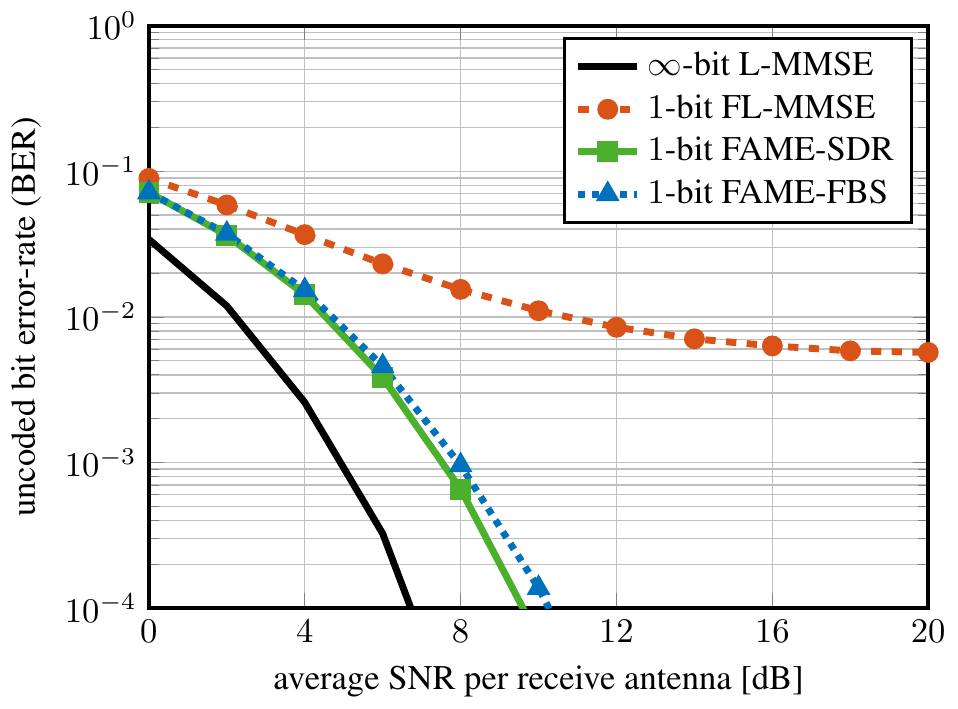}\label{fig:ber_64x4}}
\caption{Uncoded bit error-rate (BER) for two MU-MIMO systems in an i.i.d. Rayleigh-fading scenario. All equalizers, except for L-MMSE, use 1-bit finite-alphabet equalization matrices. (a) FAME-SDR approaches the performance of an exhaustive search, while being scalable for systems with more BS antennas. (b) FAME-FBS with $t_\text{max}=30$ iterations  achieves similar performance as FAME-SDR but at a significantly lower complexity.}\label{fig:ber_small}
\end{figure}

\begin{figure*}[tp]
\centering
\subfigure[i.i.d. Rayleigh]{\includegraphics[width=.64\columnwidth]{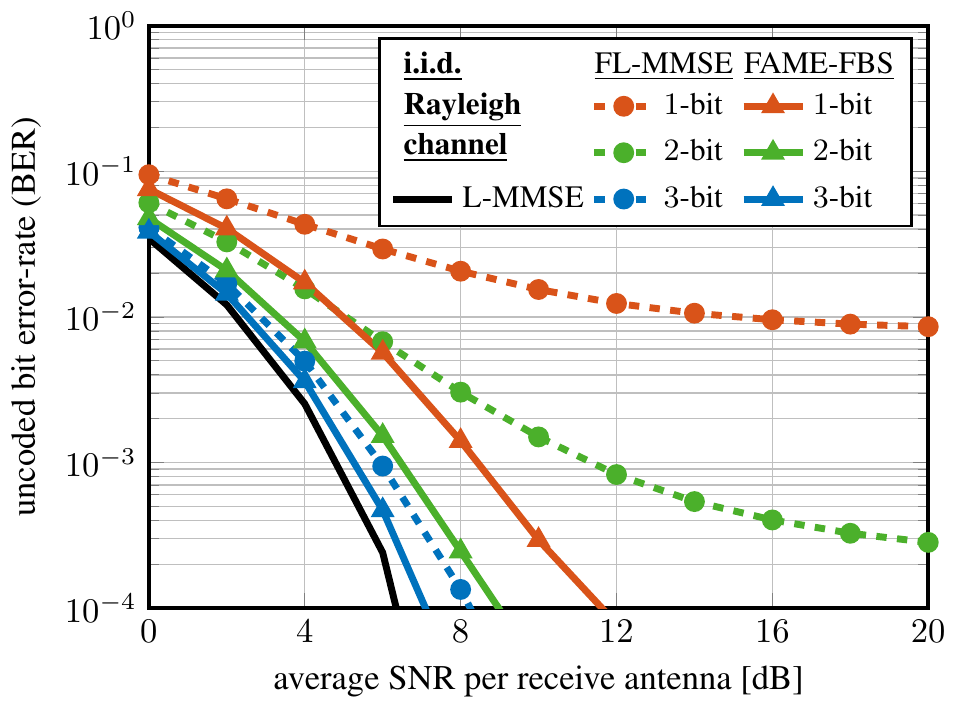}\label{fig:ber_rayleigh}}
\hfill
\subfigure[QuaDRiGa non-LoS]{\includegraphics[width=.64\columnwidth]{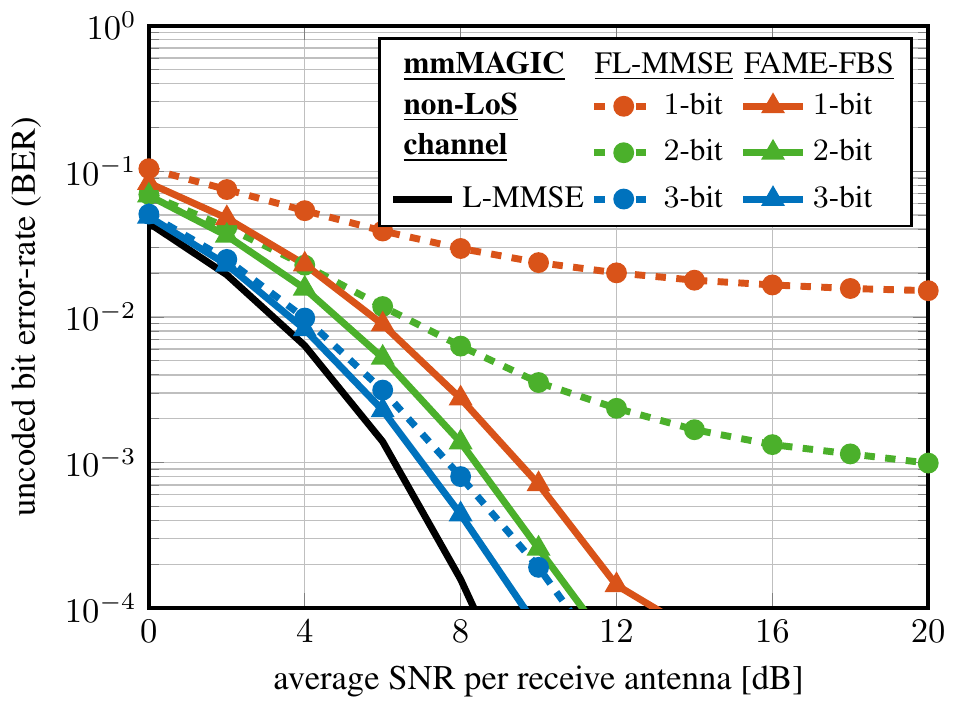}\label{fig:ber_nlos}}
\hfill
\subfigure[QuaDRiGa LoS]{\includegraphics[width=.64\columnwidth]{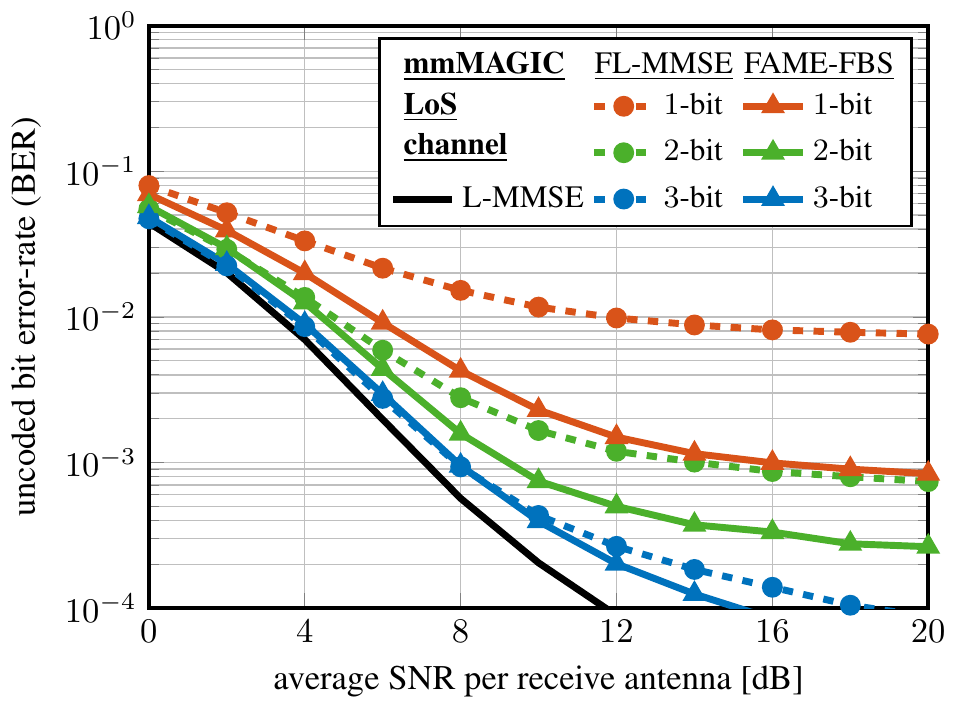}\label{fig:ber_los}}
\caption{Uncoded bit {error-rate} (BER) for a $B=256$ BS antenna, $U=16$ UE, $16$-QAM system under different channel models. FAME-FBS runs $t_\text{max}=5$ iterations for the i.i.d. Rayleigh channel and $t_\text{max}=20$ iterations for both QuaDRiga channels. FAME-FBS is initialized with the MRC equalizer $\bH^H$ for all cases, except for the $3$-bit QuaDRiGa LoS scenario, where FAME-FBS is initialized with the low-resolution part of FL-MMSE and runs for $t_\text{max}=3$ iterations. FAME-FBS outperforms FL-MMSE in all considered scenarios. The performance of finite-alphabet equalizers meets that of L-MMSE with~$6$~bits.}\label{fig:ber_all}
\end{figure*}

\begin{figure*}[tp]
\centering
\subfigure[i.i.d. Rayleigh]{\includegraphics[width=.64\columnwidth]{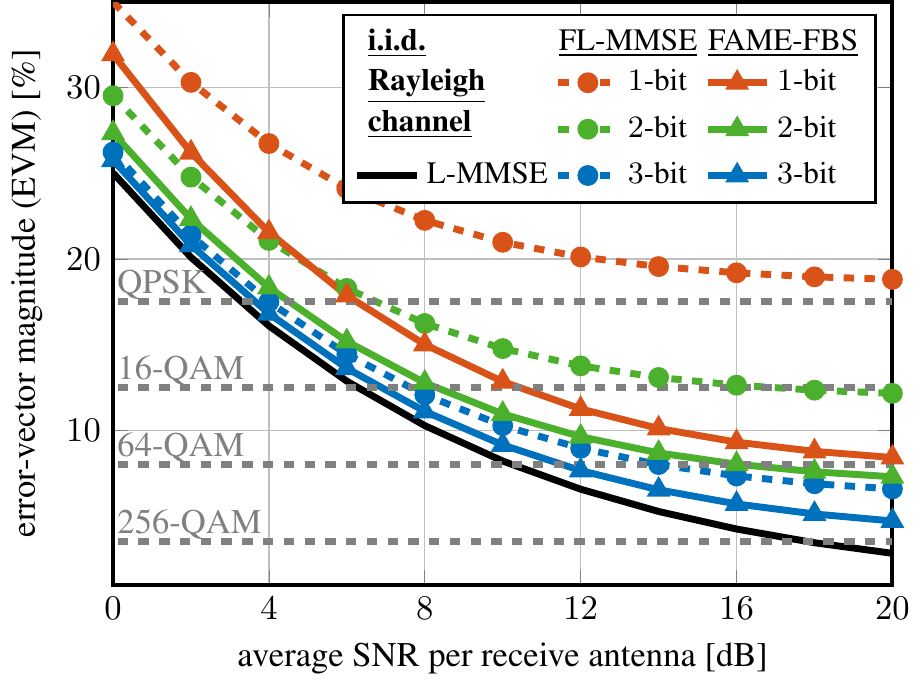}\label{fig:evm_rayleigh}}
\hfill
\subfigure[QuaDRiGa non-LoS]{\includegraphics[width=.64\columnwidth]{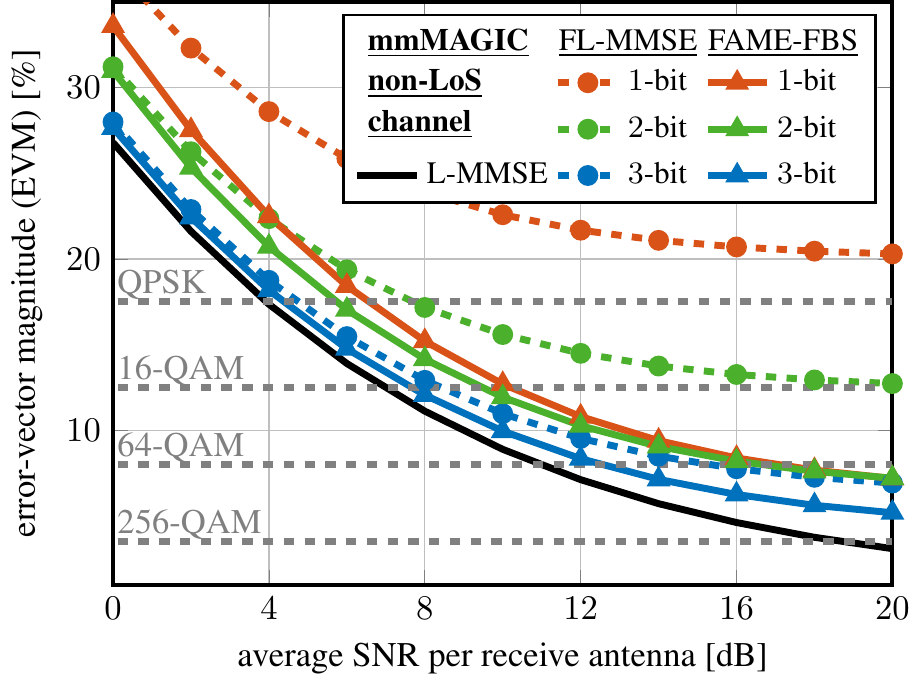}\label{fig:evm_nlos}}
\hfill
\subfigure[QuaDRiGa LoS]{\includegraphics[width=.64\columnwidth]{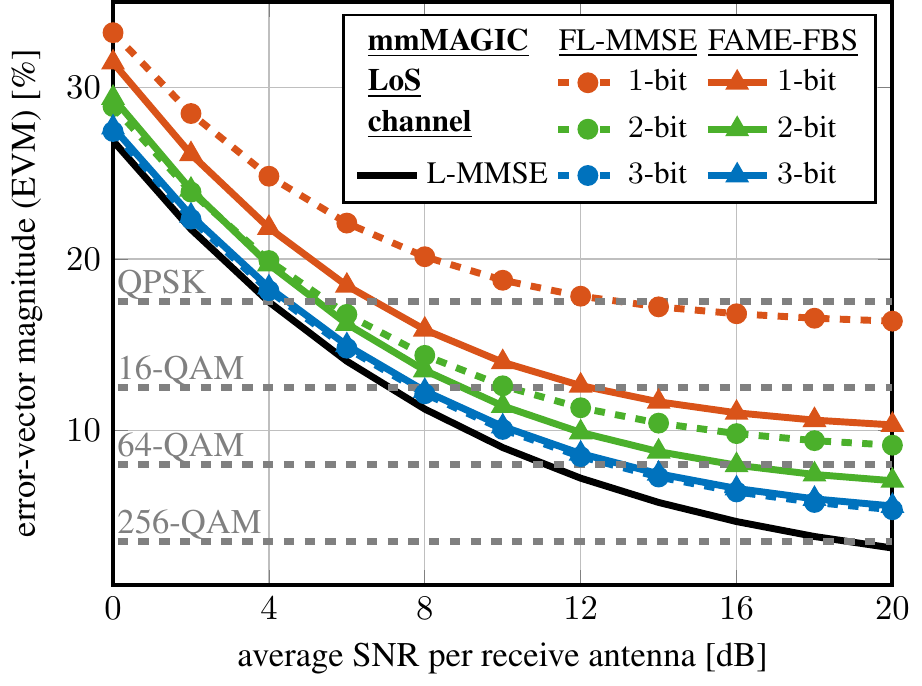}\label{fig:evm_los}}
\caption{Error-vector magnitude (EVM) for a $B=256$ BS antenna, $U=16$ UE system under different channel models. The gray dashed lines indicate the EVM requirements established by the 3GPP 5G NR  technical specification \cite{3gpp19a}. The details for FAME-FBS are given in \fref{fig:ber_all}. FAME-FBS significantly outperforms FL-MMSE in all considered scenarios when using $1$- and $2$-bit finite-alphabet equalization matrices.}\label{fig:evm_all}
\end{figure*}

\subsection{Simulation Results}
\label{sec:sim}
We now evaluate the uncoded BER performance of FAME-based algorithms, and compare it to infinite-precision L-MMSE equalization and FL-MMSE. 
The following simulation results are obtained by carrying out $10^4$ Monte-Carlo trials.
The per-iteration parameters $\{\gamma^{(t)}\}$, $\{\tau^{(t)}\}$, and $\{\nu^{(t)}\}$ of FAME-FBS are tuned using a neural network as in \cite{balatsoukas-stimming19a}; the neural network is trained using $10^4$ channel realizations, which differ from the ones used to evaluate the BER.
In practice, we have observed that $\gamma^{(t)}=1.1$ and $\nu^{(t)}=1.1$, $t=1,\ldots,t_\text{max}$, constitute good values for initializing the neural network regardless of the system configuration; good initializers for $\tau^{(t)}$ vary from $2^{-4}$ to $2^{-9}$ for the systems considered in this work.
For all equalizers, we quantize the entries of the channel matrices~$\bH$ to $8$ bits per real and imaginary components. 
In addition, the received signal vectors are quantized to $7$ bits  per real and imaginary components, which is sufficient to achieve virtually the same performance as with double-precision representation.  

In \fref{fig:ber_small}, we show uncoded BER for two different MU-MIMO systems in an i.i.d.\ Rayleigh-fading scenario.
For the $B=8$ BS antenna, $U=2$ UE system shown in \fref{fig:ber_8x2}, we see that $1$-bit FAME-EXH significantly outperforms $1$-bit FL-MMSE, which suffers from an error floor. 
Since the complexity of FAME-EXH scales exponentially in $B$, it cannot be used for significantly larger systems.
Hence, we also show the performance of FAME-SDR, which has a $2$\,dB loss at a BER of $10^{-3}$ compared to FAME-EXH.  
Since FAME-SDR scales to systems with more BS antennas, we also show a  $B=64$ BS antenna, $U=4$ UE system in \fref{fig:ber_64x4}.
In this scenario, FAME-SDR continues to substantially outperform $1$-bit FL-MMSE.
As discussed in \fref{sec:compl}, however, FAME-SDR does not scale to systems with more BS antennas, whereas FAME-FBS exhibits the same asymptotic complexity scaling as L-MMSE and FL-MMSE equalization. 
From \fref{fig:ber_small}, we see that FAME-FBS performs on par with FAME-SDR but at much lower complexity.

In \fref{fig:ber_rayleigh}, we show the BER performance of finite-alphabet equalization matrices in a $B=256$ BS antenna, $U=16$ UE system using $16$-QAM with i.i.d. Rayleigh-fading channels.
The performance behavior of $1$-bit FL-MMSE and $1$-bit FAME-FBS is similar to what we have observed for smaller systems.
The same figure also shows the performance of finite-alphabet matrices with resolutions larger than $1$ bit.
We see that the performance gap between FAME-FBS and FL-MMSE is more pronounced for $1$-bit and $2$-bit finite-alphabet equalization matrices than for $3$-bit.
We note that finite-alphabet equalizers  achieve virtually the same performance as infinite-precision L-MMSE equalization when using $6$ bits; nonetheless, $3$ bits are sufficient to operate at SNRs lower than $4$\,dB.

Since i.i.d.\ Rayleigh-fading channels are a poor model for mmWave propagation conditions, we also show the  performance of FAME-FBS in a $B=256$ BS antenna, $U=16$ UE system operating over more realistic mmWave channels generated using the QuaDRiGa model \cite{jaeckel2014quadriga}.
Concretely, in \fref{fig:ber_nlos} and \fref{fig:ber_los}, we simulate mmWave systems with a carrier frequency of $60$\,GHz within the ``mmMAGIC\_UMi'' scenario.
We consider both non-LoS (shown in \fref{fig:ber_nlos}) and LoS (shown in \fref{fig:ber_los}) propagation conditions.
We also model power control by scaling the QuaDRiGa-generated channels so that the received UE powers are in the range $\pm3$\,dB.
Specifically, for each channel realization, the UE with highest power has $4\times$ the power of the UE with the lowest power.
Furthermore, the UEs are randomly placed in a sector of $120^\circ$ in front of the BS antenna array with a distance ranging from $10$\,m to $110$\,m, and a minimum angular separation of $4^\circ$.
From \fref{fig:ber_nlos} and \fref{fig:ber_los}, we observe that FAME-FBS outperforms FL-MMSE for both non-LoS and LoS channels---essentially the same trends as for Rayleigh-fading channels.
These simulation results indicate that finite-alphabet equalization performs well under more realistic mmWave propagation conditions, while having the potential to significantly reduce power consumption and silicon area.

To further evaluate the performance of finite-alphabet equalizers, \fref{fig:evm_all} shows the EVM performance of FAME-FBS and FL-MMSE for the same system configuration and propagation conditions considered in \fref{fig:ber_all}.
The gray dashed lines indicate the EVM requirements for different modulation schemes as specified by the 3GPP 5G NR standard \cite{3gpp19a}.
\fref{fig:evm_all} confirms the trends observed in the BER simulations.
For example, while $1$-bit FL-MMSE is not able to meet the EVM requirement for QPSK in \fref{fig:evm_rayleigh} and \fref{fig:evm_nlos}, $1$-bit FAME-FBS is almost able to reach the EVM requirement for $64$-QAM.
Moreover, while FAME-FBS significantly outperforms FL-MMSE when using $1$ and $2$ bits of resolution, their EVM performance is similar for $3$ bits, in which case the performance of both finite-alphabet equalizers is close to that of infinite-precision L-MMSE.

\begin{rem}
FL-MMSE and FAME-based algorithms (FAME-EXH, FAME-SDR, and FAME-FBS) generate finite-alphabet equalization matrices as in \fref{def:finitealphabetequalizationmatrix}.
This implies that, for a fixed equalizer resolution, all the algorithms proposed in this paper produce a low-resolution matrix $\bX^H$ whose entries belong to the same finite alphabet $\setX$, as well as a set of post-equalization scaling factors $\boldsymbol\beta$, which are computed via \fref{eq:optimalscaling} once $\bX^H$ has been determined.
Even though all the algorithms use the same finite alphabet for the entries of $\bX^H$, FAME-based algorithms are able to achieve a better performance as they are (approximately) solving the FAME problem in~\fref{eq:FAMEcompact}.
\end{rem}

\begin{rem}
The improved performance of FAME-FBS over FL-MMSE comes at the cost of a higher complexity, as shown in \fref{tbl:complexitymatrix}.
Hence, there exists a performance-complexity trade-off between using FL-MMSE and FAME-FBS to generate finite-alphabet equalizers.
In addition, the complexity and performance of FAME-FBS can be further tuned via the number of iterations~$t_\text{max}$.
Finally, the equalizer resolution offers another performance-complexity trade-off: The use of more bits for the finite-alphabet equalization matrix improves the performance, but also increases the circuit's power consumption and silicon area---a trade-off we will study next.
\end{rem}
\section{Hardware-Level Evaluation}
\label{sec:vlsi}
To demonstrate the real-world benefits of finite-alphabet equalization, we now quantify the power and area savings that can be attained in comparison with conventional, high-resolution equalizers. 
\subsection{Equalizer Architectures}
\label{sec:arch}
To arrive at a fair comparison between finite-alphabet equalization and conventional, high-resolution equalizers, we implemented two equalization circuits: one for finite-alphabet equalization and one for high-resolution equalization.

The high-resolution equalizer computes a matrix-vector product between the $U\times B$ equalization matrix $\bW^H$ and the received vector $\bmy$. 
The matrix-vector product is computed in a column-by-column fashion by using a linear array of $U$ parallel multiply-accumulate (MAC) units over $B$ clock cycles.
The multipliers in the MAC units are high-resolution and take as input $10$-bit numbers from the equalization matrix $\bW^H$ and $7$-bit numbers from the received vector $\bmy$.
The accumulators in the MAC units use $18$ bits.
Finally, $9$ bits are taken from both real and imaginary accumulators as the outputs of each MAC unit.
These outputs correspond to the estimates $\hat\bms=\bW^H\bmy$.

The finite-alphabet equalizer computes a low-resolution matrix-vector product between the $U\times B$  finite-alphabet matrix~$\bX^H$ and the received vector $\bmy$.
This matrix-vector product is implemented in the same way as in the traditional equalizer, with the difference that far fewer bits are used for the multipliers and accumulators.
The multipliers take as input $r$-bit numbers from~$\bX^H$ and $7$-bit numbers from $\bmy$, while the accumulators use $r+13$ bits (except for the case where $r=1$, where the accumulators use {$13$} bits).
We take $9$ bits from the accumulators in each MAC unit as the output of the low-resolution matrix-vector product $\bX^H\bmy$.
Unlike conventional equalization, the results of the $U$-dimensional vector $\bX^H\bmy$  are scaled by the values in $\boldsymbol\beta^*$.
This scaling operation is implemented with a high-resolution multiplier that computes the product between the $9$-bit $\bmx_u^H\bmy$ and the $10$-bit scaling factor~$\beta_u^*$. The output of this multiplier is represented using 9 bits per real and imaginary components and correspond to the estimates~$\hat\bms=\bV^H\bmy$.

\subsection{Implementation Results}

\begin{table}[tp]
\renewcommand{\arraystretch}{1.1}
\begin{minipage}[c]{1\columnwidth}
    \centering
    \caption{Implementation results in 28\,nm CMOS for one equalizer instance operating in a system with  $B=256$ and $U=16$  }
    \label{tbl:implresults}
  \begin{tabular}{@{}lcccccc@{}}
  \toprule
  Equalizer resolution $r$ [bit] & $1$ & $2$ & $3$ & $4$ & $5$ & $10$\\
  \midrule  
  {Silicon area [$\text{mm}^2$]} & 0.06 & 0.08 & 0.10 & 0.14 & 0.16 & 0.26 \\
  {Clock freq. [GHz]} & 1.33 & 1.25 & 1.25 & 1.16 & 1.16 & 1.05 \\
  {Throughput [M\,vectors/s]} & 5.18 & 4.88 & 4.88 & 4.53 & 4.53 & 4.10 \\
  {Power\footnote{Extracted from stimuli-based post-layout simulations in the typical-typical process corner at $25^\circ\text{C}$ with a nominal power supply of 0.9V.} [mW]} & 18.5 & 29.2 & 38.8 & 42.6 & 51.3 & 57.1 \\
  \bottomrule
  \end{tabular}
  \end{minipage}
  \end{table}

\begin{figure}[tp]
\centering
\subfigure[Power vs. equalizer resolution]{\includegraphics[width=.48\columnwidth]{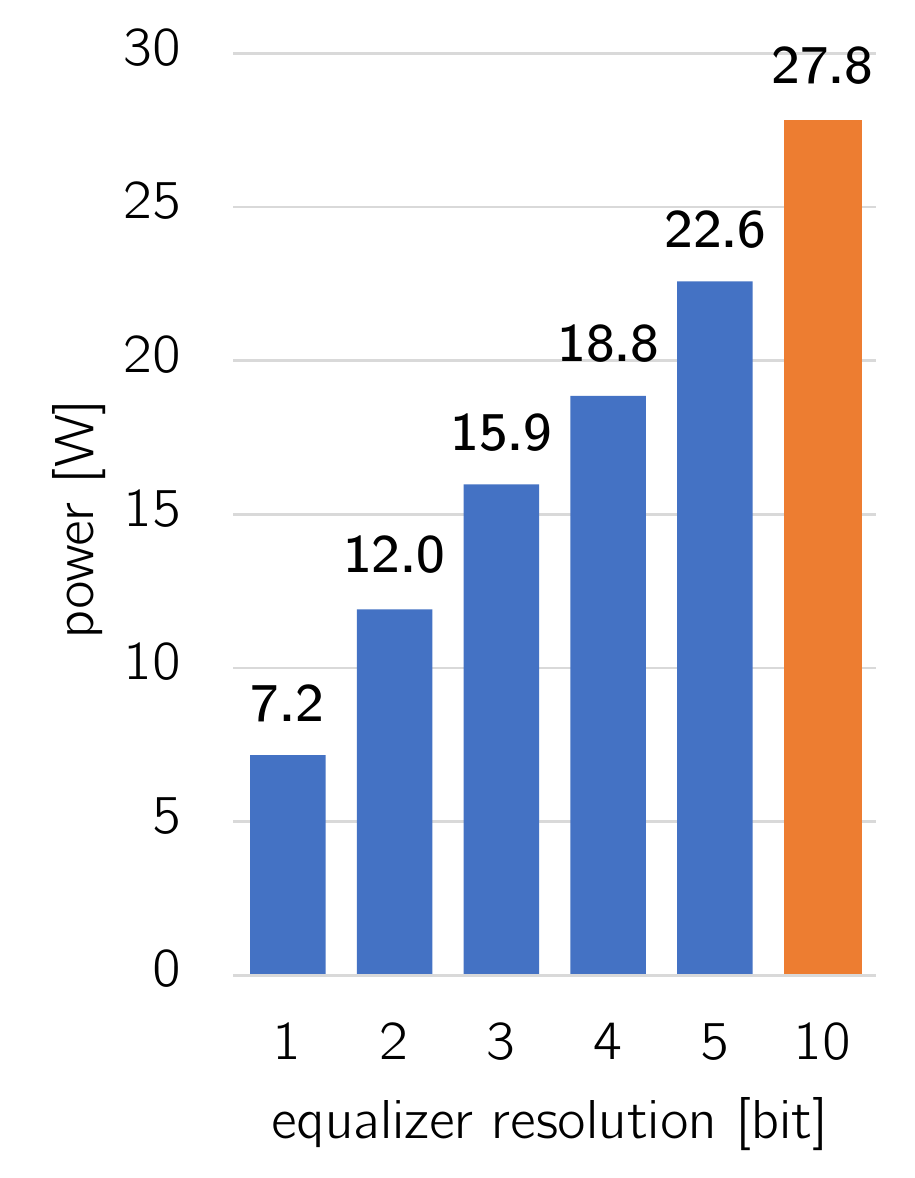}\label{fig:bits_power}}
\hfill
\subfigure[Area vs. equalizer resolution]{\includegraphics[width=.48\columnwidth]{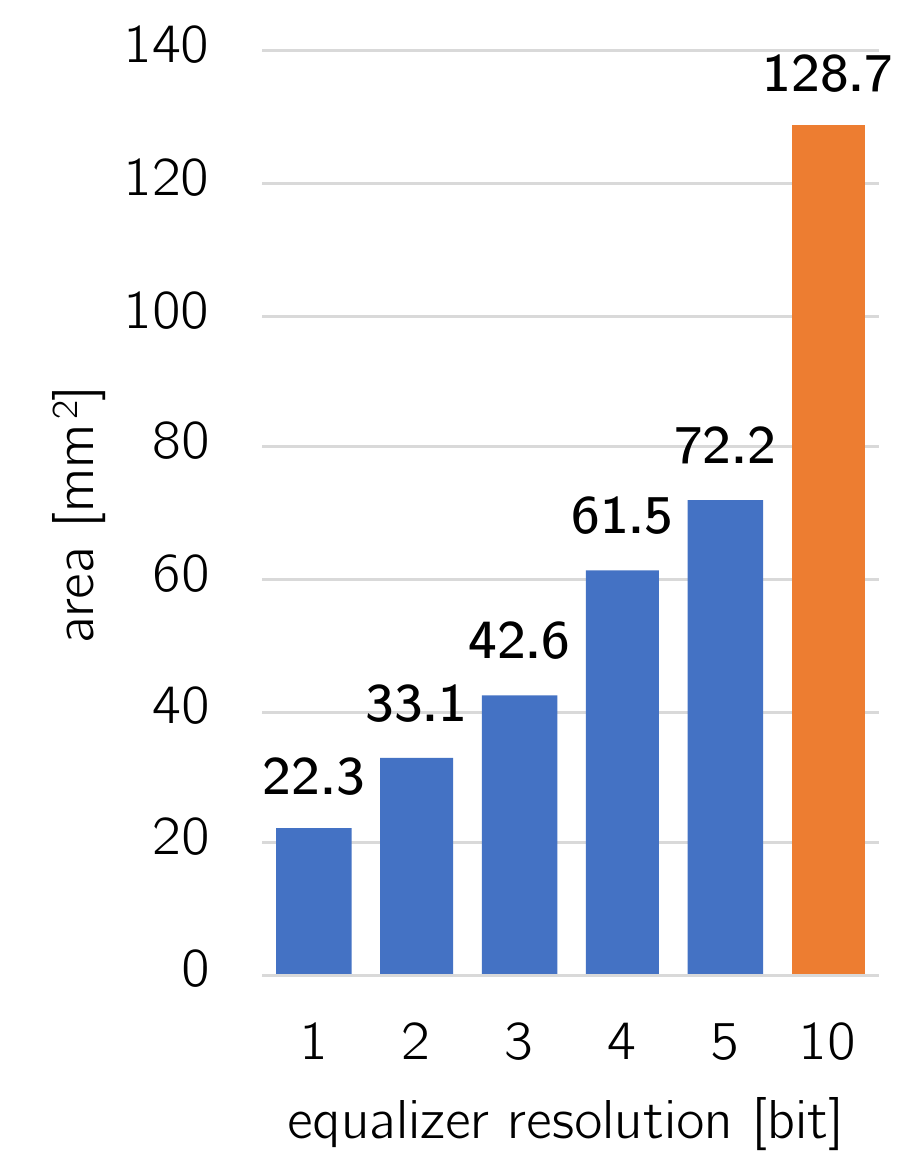}\label{fig:bits_area}}
\caption{Power and area consumed by equalizer hardware designs in $28$\,nm CMOS technology for a $B=256$ BS antenna, $U=16$ UE massive MU-MIMO system with varying equalizer resolution.
All equalizers operate at a rate of $2$ G\,vectors/s and $7$ bits are used to represent the entries of the received vector~$\bmy$. For an equalizer resolution lower than $6$ bits, we use a finite-alphabet equalizer consisting of a low-resolution matrix-vector product, followed by per-UE high-resolution scaling. The equalizer resolution of $10$~bit is executed with a high-resolution matrix-vector product only. Finite-alphabet equalization (shown in blue) can reduce the power and area of conventional, high-precision equalization (shown in orange) by a factor of $3.9\times$ and $5.8\times$, respectively.}
\label{fig:bits_all}
\end{figure}

\fref{tbl:implresults} lists post-layout implementation results for the circuits discussed in \fref{sec:arch} implemented for a $B=256$ BS antenna, $U=16$ UE system, using a $28$\,nm CMOS technology.
The traditional, high-resolution equalizer corresponds to the design with an equalization resolution $r$ of $10$ bits, whereas the finite-alphabet equalizer was implemented for $r=\{1,2,\ldots,5\}$  bits.
To allow for a fair comparison between the different equalization circuits, we consider a scenario in which all of the designs support the same throughput.
We assume a throughput of $2$ G\,(complex-valued) vectors/s, which implies that the~$2B$ ADCs at the BS run at 2\,G\,samples/s.
As we can see from \fref{tbl:implresults}, a single instance of our equalizer designs reaches throughputs of the order of M\,vectors/s, which is well below the target throughput of 2\,G\,vectors/s.
We can, however, instantiate a time-multiplexed array of equalizers that  achieve the desired throughput (at the expense of increased area). 
Assuming no overhead for this replication approach, we can estimate the total silicon area and power consumption required to perform equalization in a high-bandwidth mmWave setting;  \fref{fig:bits_all} shows the corresponding results. 

\fref{fig:bits_power} shows the power consumption reduction achieved by lowering the equalizer resolution; \fref{fig:bits_area} shows the same effect but on silicon area.
We see that halving the number of bits used for the high-resolution equalizer already introduces substantial gains of $19\%$ and $44\%$ lower power and area, respectively.
Further reducing the equalizer resolution reduces the power and area by a factor of $3.9\times$ and $5.8\times$, respectively, when a $1$-bit finite-alphabet equalizer is used.

\begin{rem}
We note that the power and area can potentially be reduced much more.
Once the number of bits in the equalization matrix has been reduced to 5~bits or below, emerging processing-in-memory architectures, such as the one proposed in \cite{castaneda19ppac}, potentially lower the area and power (additionally to the savings above) by about $2\times$ to $4\times$.
A detailed analysis of such emerging multiplier-array architectures in combination with finite-alphabet equalization is left for future~work. 
\end{rem}
\section{Conclusions}
\label{sec:conclusions}
We have proposed finite-alphabet equalization, a paradigm in which the spatial equalization matrix contains low-resolution numbers in order to enable energy- and area-efficient equalization hardware.
To achieve an error-rate performance similar to that of conventional, high-resolution equalizers, such as the L-MMSE equalizer, we have formulated the finite-alphabet MMSE equalization (FAME) problem, which minimizes the post-equalization MSE.  
We have shown that solving the FAME  problem yields significant improvements over finite-alphabet matrices that are obtained by na\"ively quantizing the L-MMSE matrix in terms of EVM, beamforming capabilities, and uncoded BER. 
Since the FAME problem is NP-hard, we have proposed approximate algorithms that trade-off  performance with complexity.
One of the proposed algorithms, FAME-FBS, achieves a performance that is on par with semidefinite relaxation while having the same asymptotic complexity scaling as L-MMSE equalization.
We have shown that FAME-FBS significantly outperforms a baseline finite-alphabet equalizer for LoS and non-LoS massive MU-MIMO mmWave channel models in terms of EVM and uncoded BER. 
In addition, our reference VLSI implementation results in $28$\,nm CMOS have demonstrated that the use of finite-alphabet equalization is able to reduce the power and area of spatial equalization by at least a  factor of $3.9\times$ and $5.8\times$, respectively.
Thus, finite-alphabet equalization is a viable solution to combat the excessively high power consumption and area of all-digital massive MU-MIMO mmWave BS designs.

There are many avenues of future work.
A theoretical convergence and performance analysis of FAME-FBS is an interesting (but difficult) open problem.
Moreover, the development of algorithms that outperform FAME-FBS and further approach the performance of FAME-EXH at low complexity is a challenging open research direction.
The finite-alphabet equalization paradigm is also applicable to downlink precoding in massive MU-MIMO mmWave systems \cite{castaneda19a} and could be used in other applications where matrix-vector products must be computed at high rates or with low power consumption---an investigation of other applications is part of ongoing research.
\appendices 

\section{Proof of \fref{lem:FAMEequivalent}}
\label{app:FAMEequivalent}

We start by deriving the expression for the optimal scaling factor $\beta_u$ given an equalization vector $\bmx_u$ and then use the resulting quantity to simplify the optimization problem. 
We first take the Wirtinger derivative of the objective function in~\fref{eq:FAME} in the complex-valued variable $\tilde\beta^*$  and set it to zero:
\begin{align}
\frac{\partial}{\partial\tilde\beta^*} \left(\|\bme_u- \bH^H\tilde\beta\tilde\bmx\|^2 + \rho \|\tilde\beta\tilde\bmx\|^2\right) & = 0 \\
-\tilde\bmx^H\bH\bme_u + \tilde\bmx^H\bH\bH^H\tilde\bmx\tilde\beta + \rho\|\tilde\bmx\|^2\tilde\beta & = 0.
\end{align}
Since $\bH\bme_u=\bmh_u$, we obtain~\fref{eq:optimalscaling} by solving for $\tilde{\beta}$. 
To obtain~\eqref{eq:FAMEcompact}, we substitute~\fref{eq:optimalscaling} into \fref{eq:FAME}
and simplify the resulting expression using algebraic manipulations. 
Concretely, we carry out the steps listed in \fref{eq:start}--\fref{eq:stop}.

\begin{figure*}[!t]
\normalsize
\setcounter{equation}{35} 

\begin{align}
& \vecnorm{\bme_u - \matH^H \beta(\tilde\bmx)\tilde\vecx }^2_2 + \rho|\beta(\tilde\vecx)|^2\|\tilde\bmx\|_2^2 \label{eq:start}\\
&\qquad  = \vecnorm{\bme_u  -   \frac{\bH^H\tilde\bmx\tilde\bmx^H\bH }{\|\bH^H\tilde\bmx\|_2^2+\rho\|\tilde\bmx\|_2^2}\bme_u  }^2_2 + \rho \frac{ |\tilde\bmx^H\bH\bme_u|^2 }{\left(\|\bH^H\tilde\bmx\|_2^2 + \rho\|\tilde\bmx\|_2^2 \right)^2} \|\tilde\bmx\|_2^2 \\
&\qquad = \vecnorm{\left( \bI - \frac{\bH^H\tilde\bmx\tilde\bmx^H\bH }{\|\bH^H\tilde\bmx\|_2^2+\rho\|\tilde\bmx\|_2^2}\right)\!\bme_u  }^2_2 + \rho \frac{ |\tilde\bmx^H\bH\bme_u|^2 }{\left(\|\bH^H\tilde\bmx\|_2^2 + \rho\|\tilde\bmx\|_2^2 \right)^2} \|\tilde\bmx\|_2^2 \\
&\qquad = \|\bme_u\|_2^2 - 2 \frac{\bme_u^H\bH^H\tilde\bmx\tilde\bmx^H\bH\bme_u }{\|\bH^H\tilde\bmx\|_2^2+\rho\|\tilde\bmx\|_2^2} +  \frac{\bme_u^H\bH^H\tilde\bmx\tilde\bmx^H\bH\bH^H\tilde\bmx\tilde\bmx^H\bH \bme_u}{\left(\|\bH^H\tilde\bmx\|_2^2+\rho\|\tilde\bmx\|_2^2\right)^2} +\rho \frac{ |\tilde\bmx^H\bH\bme_u|^2 }{\left(\|\bH^H\tilde\bmx\|_2^2 + \rho\|\tilde\bmx\|_2^2 \right)^2} \|\tilde\bmx\|_2^2 \\
&\qquad = 1 - 2\frac{|\tilde\bmx^H\bH\bme_u|^2 }{\|\bH^H\tilde\bmx\|_2^2+\rho\|\tilde\bmx\|_2^2}   +  \frac{\|\bH^H\tilde\bmx\|_2^2 |\tilde\bmx^H\bH \bme_u|^2}{\left(\|\bH^H\tilde\bmx\|_2^2+\rho\|\tilde\bmx\|_2^2\right)^2} + \rho\frac{ |\tilde\bmx^H\bH\bme_u|^2 }{\left(\|\bH^H\tilde\bmx\|_2^2 + \rho\|\tilde\bmx\|_2^2 \right)^2}\|\tilde\bmx\|_2^2 \\
&\qquad = 1 +  \frac{|\tilde\bmx^H\bH\bme_u|^2 }{\|\bH^H\tilde\bmx\|_2^2+\rho\|\tilde\bmx\|_2^2} \left( -2  +  \frac{\|\bH^H\tilde\bmx\|_2^2 }{\|\bH^H\tilde\bmx\|_2^2+\rho\|\tilde\bmx\|_2^2}  + \frac{ \rho \|\tilde\bmx\|_2^2  }{\|\bH^H\tilde\bmx\|_2^2 + \rho\|\tilde\bmx\|_2^2 } \right) \\
& \qquad= 1 +  \frac{|\tilde\bmx^H\bH\bme_u|^2 }{\|\bH^H\tilde\bmx\|_2^2+\rho\|\tilde\bmx\|_2^2} \left( -2\frac{\|\bH^H\tilde\bmx\|_2^2+\rho\|\tilde\bmx\|_2^2 }{\|\bH^H\tilde\bmx\|_2^2+\rho\|\tilde\bmx\|_2^2}   +  \frac{\|\bH^H\tilde\bmx\|_2^2 }{\|\bH^H\tilde\bmx\|_2^2+\rho\|\tilde\bmx\|_2^2}  + \frac{ \rho \|\tilde\bmx\|_2^2  }{\|\bH^H\tilde\bmx\|_2^2 + \rho\|\tilde\bmx\|_2^2 } \right) \\
& \qquad= 1 -  \frac{|\tilde\bmx^H\bH\bme_u|^2 }{\|\bH^H\tilde\bmx\|_2^2+\rho\|\tilde\bmx\|_2^2}  =  1 -  \frac{|\bmh^H_u\tilde\bmx|^2 }{\|\bH^H\tilde\bmx\|_2^2+\rho\|\tilde\bmx\|_2^2}. \label{eq:stop}
\end{align}

\hrulefill
\vspace*{2pt}
\end{figure*}

\section{Complexity Counts for Computing Equalization Matrices with Different Algorithms}
\label{app:count}

In what follows, we ignore the complexity of reciprocals, square roots, and additions. The numbers in parentheses are real-valued multiplications, where we assume that a complex-valued multiplication requires four real-valued multiplications. 

\subsection{Complexity of Explicit L-MMSE}
\label{app:compl_lmmse}
Explicit L-MMSE equalization corresponds to computing $\bW^H=(\rho\bI_U+\bH^H\bH)^{-1}\bH^H$, which can be achieved at low complexity using the approach detailed in \cite{studer2011asic}.
First, we calculate $\bA=\rho\bI_U+\bH^H\bH$ ($2BU^2$).
Then, we apply a Cholesky decomposition to $\bA$ so that $\bA=\bL\bL^H$ ($\frac{2}{3}U^3-\frac{2}{3}U$).
Next, we compute the inverse of $\bL$ via back-substitution ($\frac{2}{3}U^3-\frac{5}{3}U+1$), to calculate $\bA^{-1}=\bL^{-H}\bL^{-1}$ ($\frac{2}{3}U^3+\frac{1}{3}U$).
Finally, we obtain $\bW^H=\bA^{-1}\bH^H$ ($4BU^2-2BU$).
The total complexity of the explicit L-MMSE equalizer is $2U^3+6BU^2-2BU-2U+1$.
Since in massive MU-MIMO systems we typically have $B\gg U$, the asymptotic complexity scales as $O(BU^2)$.

\subsection{Complexity of FL-MMSE}
\label{app:compl_qlmmse}
We start by computing the explicit L-MMSE equalizer, which, as shown in \fref{app:compl_lmmse}, entails a complexity of $2U^3+6BU^2-2BU-2U+1$.
Then, we quantize the entries of the L-MMSE equalizer. We will not count the complexity of quantization as there are hardware-efficient ways to do so.
Now that $\bX^H$ has been determined, the next step is to compute the optimal scaling factor $\beta_u(\bmx_u)$ for each UE.
We need to first calculate $\bH^H\bmx_u$ ($4BU$) from which we also extract $\bmx_u^H\bmh_u$.
Then, we compute the $\ell_2$-norm of $\bH^H\bmx_u$ and $\bmx_u$ ($2U$ and $2B$, respectively).
The next steps are to scale $\|\bmx_u\|^2_2$ by $\rho$ ($1$ multiplication), and obtain $\beta_u(\bmx_u)$ by multiplying $\bmx_u^H\bmh_u$ and the multiplicative inverse of $\|\bH^H\bmx_u\|_2^2+\rho\|\bmx_u\|^2_2$ ($2$ multiplications).
Then, computing $\beta_u(\bmx_u)$ for all UEs $u=1,\ldots,U$ has a complexity of $4BU^2+2BU+2U^2+3U$.
Thus, computing $\bX^H$ and $\boldsymbol\beta$ for FL-MMSE equalization has a total complexity of $10BU^2+2U^3+2U^2+U+1$.
As expected, FL-MMSE has the same asymptotic complexity scaling  $O(BU^2)$ as the L-MMSE equalizer.

\subsection{Complexity of FAME-FBS}
For each UE $u=1,\dots,U$, one instance of FAME-FBS is executed. Each FAME-FBS instance requires an iterative procedure with $t_\text{max}$ iterations followed by the computation of $\beta_u(\bmx_u)$.
In the iterative procedure, we compute $\bH^H\tilde\bmx$ ($4BU$) and scale one of its entries with $\gamma^{(t)}$ ($2$ multiplications). 
With this, we have computed the vector ($\bH^H-\gamma^{(t)}\bme_u\bme_u^H\bH^H)\tilde\bmx$, which we now multiply with $\bH$ ($4BU$) to obtain the gradient $\nabla f(\tilde\bmx)$ in \fref{eq:fbs_grad}.
The next step in FAME-FBS is to scale $\nabla f(\tilde\bmx)$ with $\tau^{(t)}$ ($2B$) to compute $\tilde\bmz^{(t+1)}$ in \fref{eq:fbs_z}.
Then, the entries of $\tilde\bmz^{(t+1)}$ are scaled by $\nu^{(t)}$ ($2B$), completing one FAME-FBS iteration.
Hence, to compute $\bmx_u$, FAME-FBS requires $(8BU+4B+2)t_\text{max}$ real-valued products.
As shown in \fref{app:compl_qlmmse}, computing the optimal scaling factor $\beta_u(\bmx_u)$ requires $4BU+2B+2U+3$ real-valued multiplications per UE.
Thus, the total computational complexity of FAME-FBS is  $(8t_\text{max}+4)BU^2+2U^2+2(2t_\text{max}+1)BU+(2t_\text{max}+3)U$. Hence,  FAME-FBS has an asymptotic complexity of $O(BU^2)$.

\balance


\balance

\end{document}